\documentclass[%
superscriptaddress,
nofootinbib,
 amsmath,amssymb,
 aps,
pra,
floatfix,
]{revtex4-2}
\usepackage{array,multirow,graphicx,makecell}
\usepackage[font=small,labelfont=bf,justification=justified]{caption}
\usepackage{subcaption}
\usepackage{ftnxtra}
\usepackage{booktabs}
\usepackage{dcolumn}
\usepackage{hhline}
\usepackage{fontawesome}
\usepackage{bm}
\usepackage[colorlinks,citecolor=blue,urlcolor=blue,linkcolor=blue]{hyperref}
\usepackage{color}
\usepackage{orcidlink}  % For ORCID icon and link
\usepackage{graphicx}
\usepackage{physics}
\usepackage{tabularray}
\usepackage{threeparttable}
\usepackage{comment}
\usepackage{float}
\usepackage{bm}
\usepackage{natbib}
\graphicspath{{./}{FIG_2/}}

\usepackage{xcolor}

\begin{document}

%\preprint{APS/123-QED}

%%%%%%%%%%%%%%%%%%%%%%%%%%%%%%%%%%%%%%%%%%%%%%%%%%%%%%%%%%%%%%%
\title{A Six-Parameter Teleparallel Cosmology Beyond $\Lambda$CDM: Sign-Changing Torsional Dark Energy and Implications for $H_0$}% Force line breaks with \\
%\thanks{A footnote to the article title}%

\author{Mohamed Bedair\,\orcidlink{0000-0001-6446-1613}}\email{mohamedbedair@du.edu.eg}
\affiliation{Department of Mathematics, Faculty of Science, Damietta University, Egypt}
\author{Mahmoud Hashim\,\orcidlink{0000-0003-0691-9264}}\email{mahmoud.hashim@bue.edu.eg}
\affiliation{Centre for Theoretical Physics, The British University in Egypt, P.O. Box 43, El Sherouk City, Cairo 11837, Egypt}
\author{Jackson Levi Said\,\orcidlink{0000-0002-7835-4365}}\email{jsaid01@um.edu.mt}
\affiliation{Institute of Space Sciences and Astronomy, University of Malta, Malta, MSD 2080 Department of Physics, University of Malta, Malta}
\author{Waleed El Hanafy\,\orcidlink{0000-0002-0097-6412}}\email{waleed.elhanafy@bue.edu.eg}
\affiliation{Centre for Theoretical Physics, The British University in Egypt, P.O. Box 43, El Sherouk City, Cairo 11837, Egypt}
%\altaffiliation[Also at ]{add alternative affiliation}%Lines break automatically or can be forced with \\

%\date{\today}% It is always \today, today,
             %  but any date may be explicitly specified

\begin{abstract}
We investigate a particular case of the extended exponential infrared $f(T)$ teleparallel gravity, in which the geometric sector naturally produces an \emph{effective} dark--energy density that evolves from negative values in the past to positive values at late times. This behaviour could be motivated by observational results providing a compelling motivation for a geometric, sign--changing dark--energy scenario within modified gravity. We demonstrate that the parameter space of the present model contains only six parameters similar to $\Lambda$CDM. A Markov Chain Monte Carlo (MCMC) analysis using \textit{Planck}, DESI, and Type~Ia supernova data yields a well--constrained transition redshift of $z_{tr} \gtrsim 1.62$, accompanied by a transition in the effective equation of state to the phantom regime ($w<-1$) for $z<z_{tr}$. Since the effective dark energy originates from modified geometric degrees of freedom, no instabilities or violations of energy conditions arise. The model naturally accounts for the $H_0$ tension, where Planck+DESI data combination gives $H_0= 72.13 \pm 0.28~{\rm km}\,{\rm s}^{-1}\,{\rm Mpc}^{-1}$ in a better agreement with local measurements than $\Lambda$CDM which gives $H_0=68.46 \pm 0.30~{\rm km}\,{\rm s}^{-1}\,{\rm Mpc}^{-1}$. However, the model is disfavored in comparison to $\Lambda$CDM in terms of the values of $\chi^2$ of the bestfit parameters. We discuss the result among other issues related to CMB-BAO tension.
\end{abstract}

%\keywords{Suggested keywords}%Use showkeys class option if keyword
                              %display desired
\maketitle

%\tableofcontents

%%%%%%%%%%%%%%%%%%%%%%%%%%%%%%%%%%%%%%%%%%%%%%%%%%%%%%%%%%%%%%%%%%%%%%%%%%%%%%%%%%%%%%%%%%%%%%%%%%%%%%%%%%%%%%%%%%%%%%%%%%%%%%%%%%%%%%
\section{Introduction}
%%%%%%%%%%%%%%%%%%%%%%%%%%%%%%%%%%%%%%%%%%%%%%%%%%%%%%%%%%%%%%%%%%%%%%%%%%%%%%%%%%%%%%%%%%%%%%%%%%%%%%%%%%%%%%%%%%%%%%%%%%%%%%%%%%%%%%

$\Lambda$CDM concordance cosmology remains one of the most successful descriptions of the evolution of the Universe with decades of accurate predictions at astrophysical and cosmological physical scales~\cite{Peebles:2002gy,Copeland:2006wr}. On the other hand, there are strengthening observational and theoretical challenges to this framework of cold dark matter (CDM)~\cite{Baudis:2016qwx,XENON:2018voc}, a cosmological constant ($\Lambda$) realisation of dark energy~\cite{Riess:1998cb,Perlmutter:1998np}, and gravity through general relativity (GR)~\cite{Misner:1973prb,Weinberg:1972kfs}. There have been theoretical and observational challenges to the concordance model since its establishment with foundational issues in the description of the cosmological constant~\cite{Weinberg:1988cp} to the question of ultra-violet completeness~\cite{Addazi:2021xuf}, and the potential of direct observations of CDM~\cite{LUX:2016ggv,Gaitskell:2004gd}, in addition to numerous other challenges~\cite{DiValentino:2025sru}. In recent years, the level of precision in cosmological parameters has drastically increased in the local Universe~\citep{Riess:2021jrx,H0DN:2025lyy} producing results that are statistically in disharmony with predictions based on early-Universe observations. A crucial factor is that these early-Universe predictions are based directly on the $\Lambda$CDM model, which may require new physics to resolve~\cite{ACT:2023kun,Schoneberg:2022ggi}.

The disparity between the values of certain cosmological parameters can be illustrated through the Hubble constant ($H_0$) which gives a measure of the present cosmic expansion. The contrast can be illustrated through direct constraints on $H_0$ using DESI Collaboration data~\cite{Said:2024pwm} and independent constraints on the distance to the Coma cluster, giving a value of $76.5 \pm 2.2~{\rm km}\,{\rm s}^{-1}\,{\rm Mpc}^{-1}$~\cite{2025ApJ...979L...9S}, while predictions based on CMB data from the Atacama Cosmology Telescope (DR6) gives the Hubble constant, $H_0^{\rm ACT} = 68.3\pm 1.1~{\rm km}\,{\rm s}^{-1}\,{\rm Mpc}^{-1}$~\cite{ACT:2023kun} which is consistent with other CMB constraints. These measurements are reinforced by a myriad of measurements in the late-Universe~\cite{H0DN:2025lyy} that depend on independent sources, as well as different methods in the early-Universe that also use Big Bang nucleosynthesis data in combination with other observational information~\cite{Schoneberg:2022ggi} which still assume a $\Lambda$CDM fiducial model. The statistical disparity between these values has prompted a reevaluation of possible new underlying physics drivers in the evolution of the Universe.

There has been a wide range of different potential solutions suggested~\cite{CANTATA:2021ktz,Capozziello:2002rd,Capozziello:2011et,Clifton:2011jh,Nojiri:2010wj,Nojiri:2017ncd} which aim to resolve this open question. By and large, these models propose extensions to $\Lambda$CDM assuming Riemannian geometry, while non-Riemannian geometries remained poorly studied and a rich area of potential model-building. A key development in this area of gravitation is the development of teleparallel gravity (TG)~\cite{Bahamonde:2021gfp,Aldrovandi:2013wha,Cai:2015emx,Krssak:2018ywd} where the curvature-based Levi-Civita connection $\bar{\Gamma}^{\sigma}{}_{\mu\nu}$ (over-bars denote quantities based on the Levi-Civita connection in this work) with the torsion-based teleparallel connection $\Gamma^{\sigma}{}_{\mu\nu}$. In recent years, TG has gained momentum as a viable description of gravitation, and a rich area on which to build new cosmological models. The development of Riemannian geometries has been much more exhaustive while TG provides a much more intuitive approach to constructing gravitational models.

In TG, the gravitational connection expresses the geometric torsion while also having curvature-free and non-metricity properties~\cite{Aldrovandi:2013wha}. Interestingly, by combining the constituents of TG in a particular way, there exists a teleparallel equivalent of General Relativity (TEGR) that is dynamically equivalent to GR at the level of the field equations~\cite{Bahamonde:2021gfp}. In this limit, all classical predictions of GR and TEGR will be identical~\cite{Capozziello:2022zzh}. This is achieved by having a Lagrangian that is linear in the torsion scalar $T$. Taking the same approach as in $f(R)$ gravity~\cite{Capozziello:2011et,Sotiriou:2008rp,DeFelice:2010aj}, an $f(T)$ gravity~\cite{Ferraro:2006jd,Ferraro:2008ey,Bengochea:2008gz,Linder:2010py,Chen:2010va,Basilakos:2013rua,Cai:2015emx,Bahamonde:2019zea,Paliathanasis:2017htk,Farrugia:2020fcu,Bahamonde:2021srr,Bahamonde:2020bbc} can be produced by directly generalizing the TEGR action. In contrast to $f(R)$ gravity, $f(T)$ gravity results in generally second order equations of motion making the overall theory more attractive.

Building on this background, we consider an intriguing expression of $f(T)$ cosmology in which TEGR is enriched by terms that couple to produce a sign-switching effect in the late-Universe. This has shown great promise in recent years~\cite{Akarsu:2021fol} in tackling the open issue of cosmic tensions. Of particular interest for our case is that we consider a cosmology that exhibits the same number of free parameters as in $\Lambda$CDM which will have an important impact on the statistical preference of the model. In our work, we start by reviewing the fundamental physics associated with $f(T)$ gravity in Sec.~\ref{fid} where the background and perturbative cosmology is then developed. This latter aspects are crucial to producing accurate predictions for the full range of data sets that contribute to the appearance of cosmological tensions.

Following this review analysis, we discuss how we developed our new proposal for $f(T)$ gravity where the well-known exponent extension of TEGR is augmented with a term that acts as a sign-switching contributor. Here, we also discuss how the free parameters of the underlying system turn out to be equal to that of $\Lambda$CDM which gives this scenario a statistical advantage. We then establish the model's viability at background level through an analysis of its phase portrait in Sec.~\ref{phase}. In Sec.~\ref{data}, we discuss the early- and late-time data sets we consider in the work. These are based on an upcoming study by the CosmoVerse network~\cite{DiValentino:2025sru} called the Cosmology Compilation Group which seeks to provide a model agnostic assessment of different cosmological models, and will be released later in the year. Our numerical results are contained in Sec.~\ref{res} where we show the model parameter constraints, as well as the effects of this model in tackling the scalar spectral index tensions in TG and its results in the growth of large scale structure information. Finally, we close in Sec.~\ref{sec:Conclu} with some final remarks and some possible future work.

%%%%%%%%%%%%%%%%%%%%%%%%%%%%%%%%%%%%%%%%%%%%%%%%%%%%%%%%%%%%%%%%%%%%%%%%%%%%%%%%%%%%%%%%%%%%%%%%%%%%%%%%%%%%%%%%%%%%%%%%%%%%%%%%%%%%%%
 \section{\texorpdfstring{$f(T)$}{} Theory}\label{fid}
%%%%%%%%%%%%%%%%%%%%%%%%%%%%%%%%%%%%%%%%%%%%%%%%%%%%%%%%%%%%%%%%%%%%%%%%%%%%%%%%%%%%%%%%%%%%%%%%%%%%%%%%%%%%%%%%%%%%%%%%%%%%%%%%%%%%%%
In this section, we briefly discuss Weitzenb\"ock geometry, where the torsion tensor measures the deviation from Minkowski space. Also, we write the field equations when $f(T)$ function is involved in Einstein-Hilbert action and the corresponding modifications of Friedmann equations on cosmology background equations and the linear perturbation as well.

%%%%%%%%%%%%%%%%%%%%%%%%%%%%%%%%%%%%%%%%%%%%%%%%%%%%
\subsection{Mathematical Structure}
%%%%%%%%%%%%%%%%%%%%%%%%%%%%%%%%%%%%%%%%%%%%%%%%%%%%
We consider spacetime as a four-dimensional manifold and introduce four independent orthonormal vector fields at every point, called a \emph{tetrad}. Denoting the metric tensor by $g_{\mu\nu}$, the tetrad satisfies
\[
e{^a}{_\mu} e{^b}{_\nu} g^{\mu\nu} = \eta^{ab}, \qquad
e{^a}{_\mu} e{^b}{_\nu} \eta_{ab} = g_{\mu\nu}.
\]
where $\eta^{ab}$ denotes the Minkowski metric of the tangent space in the tetrad basis,
which takes the form $\mathrm{diag}(-1,1,1,1)$ in the $(-,+,+,+)$ signature.
In the teleparallel framework, the tetrad is the fundamental quantity and has $16$ independent components. This is more than in General Relativity (GR) because local Lorentz invariance is non-trivial. We define a quantity $\omega$, called the spin connection, which is simply the connection associated with the tetrad. It is defined by~\cite{DeAndrade:2000sf,2013tgif.book.....A,Krssak:2015oua,Krssak:2018ywd}
\begin{equation}\label{eq:spin}
   \nabla_\mu e{_a}{^\alpha}=\omega{^\alpha}{_{a\mu}}.
\end{equation}
Using this definition, the affine connection $\Gamma^\alpha{_{\mu\nu}}$ can be written as
\begin{equation}\label{eq:conn}
    \Gamma{^\alpha}_{\mu\nu}
    =
    e{_a}{^\alpha}\partial_\nu e{^a}{_\mu}
    +
    e^{a}{_\mu}\omega{^\alpha}_{a\nu}.
\end{equation}
Within the framework of teleparallel gravity, we choose a flat, metric-compatible connection. In this case, it is possible to choose a tetrad such that the spin connection vanishes,
$\omega{^\alpha}{_{a\mu}}=0$. Substituting this condition into Eq.~(\ref{eq:conn}) yields the Weitzenb\"{o}ck connection,
\begin{equation}
    \Gamma{^\alpha}_{\mu\nu}=e{_a}{^\alpha}\partial_\nu e{^a}{_\mu}.
\end{equation}
It should be noted that this form of the connection corresponds to a specific choice of tetrad. Consequently, any equations constructed directly from it are, in general, not covariant under local Lorentz transformations.

In this formulation, gravity is no longer described by spacetime curvature but by torsion, defined as
\[
T^\alpha_{\;\;\mu\nu} = e{_a}{^\alpha} \partial_\mu e{^a}{_\nu}-e{_a}^\alpha \partial_\nu e{^a}{_\mu}=\Gamma{^\alpha}_{\nu\mu} - \Gamma{^\alpha}_{\mu\nu},
\]
which is an antisymmetric tensor in the last two indices. Then, the teleparallel torsion scalar can be defined as
\[
T = \frac{1}{4} T^{\rho\mu\nu} T_{\rho\mu\nu}
    + \frac{1}{2} T^{\rho\mu\nu} T_{\nu\mu\rho}
    - T_{\mu}T^{\mu},
\]
where $T_{\mu}=T^{\alpha}_{\;  \alpha \mu}=-T^{\alpha}_{\;   \mu \alpha}$. Once the metric tensor is defined, the Levi-Civita connection $\bar{\Gamma}^{\alpha}_{\beta \gamma}$ can be obtained, where
\[
\bar{\Gamma}^{\alpha}_{\beta \gamma}=\frac{1}{2}g^{\alpha \sigma}(\partial_{\beta}g_{\gamma \sigma}+\partial_{\gamma}g_{\beta \sigma}-\partial_{\sigma}g_{\gamma \beta}).
\]
The difference between Weitzenb\"{o}ck and Levi-Civita connections defines the contortion tensor $K^\alpha_{\;\;\beta\gamma}$, which can be expressed in terms of the torsion tensor as follows
\[
K^\alpha_{\;\;\beta\gamma} = \frac{1}{2} \left(
T_\beta{}^\alpha{}_\gamma + T_\gamma{}^\alpha{}_\beta - T^\alpha_{\;\;\beta\gamma}
\right).
\]

It can be shown that the torsion scalar differs from the curvature scalar derived from the Levi-Civita connection by a total derivative:
\[
R(\bar{\Gamma}) = -T + 2 \nabla_\mu T^\mu.
\]

Thus, an action based on $T$ is equivalent to GR. To go beyond GR, we generalize the action to be a non-linear function of $T$:
\[
S =  \int d^4x \,e\left(\frac{-1}{2\kappa^2} f(T)+L_m \right),
\]
where $e = \det(e_\mu^a)$ and $L_m$ is the Lagrangian of the matter field and the factor $\kappa$ define from Newton's gravitational constant by $\kappa^2=\frac{8\pi G}{c^4}$. Variation of this action with respect to the tetrad leads to field equations that are different from GR and are generally non-linear in $T$. Note that this theory is not trivially Lorentz invariant, so the field equations are not symmetric in general.

Variation of the action with respect to tetrad gives~\cite{Li:2011wu}
\begin{equation}
f_T\mathrm{G_{\mu\nu}} - \frac{1}{2}g_{\mu\nu}(Tf_T-f) + S_{\mu\nu}^{\; \; \; \sigma} \partial_{\sigma}Tf_{TT}=\mathcal{\kappa}^2\mathcal{T_{\mu\nu}}
\label{fid_eq}
\end{equation}
with  $\mathrm{G_{\mu\nu}}$ is Einstein tensor, and the superpotential $S_{\mu\nu}^{\; \; \; \sigma}$
\begin{equation}
S_\rho^{\;\;\mu\nu} =   K^{\mu\nu}_{\;\;\;\;\rho} + \delta^\mu_\rho \, T^{\alpha \nu}_{\;\;\;\;\alpha} - \delta^\nu_\rho \, T^{\alpha \mu}_{\;\;\;\;\alpha} ,
\end{equation}
and the matter energy-momentum tensor
\begin{equation}
\mathcal{T}_\rho^{\;\;\nu} = -\frac{1}{e} e^{a\nu}\frac{\delta(e L_m)}{\delta e^\rho_{\;a}}.
\end{equation}
The energy-momentum tensor $\mathcal{T}^{\mu\nu}$ satisfies the covariant conservation equation
\begin{equation}
    \nabla_{\mu}\mathcal{T}^{\mu \nu}=0 \, .
\end{equation}

For a perfect fluid, the energy--momentum tensor is defined as
\begin{equation}
    \mathcal{T}_{\mu \nu}
    =
    \rho\, u_{\mu} u_{\nu}
    +
    p \left( g_{\mu \nu} + u_{\mu} u_{\nu} \right),
\end{equation}
where $u_{\mu}$ denotes the four-velocity of the fluid, and $\rho$ and $p$ represent the energy density and pressure, respectively.

%%%%%%%%%%%%%%%%%%%%%%%%%%%%%%%%%%%%%%%%%%%%%%%%%%%%
\subsection{Background Cosmology}\label{sec:bkg}
%%%%%%%%%%%%%%%%%%%%%%%%%%%%%%%%%%%%%%%%%%%%%%%%%%%%
Assuming a homogeneous and isotropic background, the flat FLRW metric in $c=1$ units is
\[
ds^2 = -dt^2 + a^2(t) \delta_{ij} dx^i dx^j,
\]
where $a(t)$ is the scale factor. A tetrad reproducing this metric is\footnote{We note that this tetrad choice leads to a consistent $f(T)$ configuration without involving nonphysical degrees of freedom due to local Lorentz invariance~\cite{DeAndrade:2000sf,2013tgif.book.....A,Krssak:2015oua,Krssak:2018ywd}.}
\[
e_a{^\mu} = \mathrm{diag}(1, a(t), a(t), a(t)).
\]
Using this tetrad in the field equations~\eqref{fid_eq}, we obtain
\begin{equation}
3H^2 = \kappa^2 (\rho_m + \rho_r) + \frac{f}{2} + \frac{T}{2} - T f_T,
\label{fir_frid}
\end{equation}
\begin{equation}
\dot{H} = \frac{-\kappa^2}{2} \frac{\rho_m + p_m + \rho_r + p_r}{f_T + 2 T f_{TT}},
\label{scnd_frid}
\end{equation}
where $H = \dot{a}/a$ is the Hubble parameter and
\begin{equation}\label{eq:Tsc}
    T = 6 H^2.
\end{equation}

In the late universe, when matter dominates radiation, we can write
\begin{equation}\label{eq:newexp_Fr_eqns}
3H^2 = \kappa^2 (\rho_m + \rho_{DE}), \qquad
\dot{H} = \frac{-\kappa^2}{2} (\rho_m + p_m + \rho_{DE} + p_{DE}),
\end{equation}
with
\begin{equation}\label{eq:newexp_tor}
    \rho_{DE} = \frac{1}{2\kappa^2} (-2 T f_T + T + f), \qquad
p_{DE} = \frac{1}{2\kappa^2} \frac{-f + T f_T - 2 T^2 f_{TT}}{f_T + 2 T f_{TT}}.
\end{equation}
It should be understood that the ``\textit{dark energy}" density and pressure in the above are generated by geometric sources, i.e. $f(T)$ teleparallel gravity, and it cannot be considered as physical dark energy fluid. It is obvious that $\rho_{DE}=0$ and $p_{DE}=0$ where $f(T)=T$ which recovers the GR theory as a limiting case. We refer to $f(T)$ teleparallel gravity as $f$CDM, where $f(T)$ counterpart plays the role of dark energy explaining late accelerated expansion as a modified gravity. In this sense, we define the torsional dark energy equation of state is
\begin{equation}
w_{DE} = \frac{p_{DE}}{\rho_{DE}} = -1 + \frac{(-f + 2 T f_T)(f_T + 2 T f_{TT} - 1)}{(-f - T + 2 T f_T)(f_T + 2 T f_{TT})},
\end{equation}
while the effective (total) equation of state is
    \[w_\mathrm{eff} = -1 - \frac{2 \dot{H}}{3 H^2}.\]
In addition, the continuity equation for each component is
\begin{equation}\label{eq:cont}
\dot{\rho}_i + 3 H (1 + w_i) \rho_i = 0, \quad i = r, m, DE.
\end{equation}

%%%%%%%%%%%%%%%%%%%%%%%%%%%%%%%%%%%%%%%%%%%%%%%%%%%%
\subsection{Linear Perturbations in Cosmology}
%%%%%%%%%%%%%%%%%%%%%%%%%%%%%%%%%%%%%%%%%%%%%%%%%%%%

We introduce linear scalar perturbations in the Newtonian gauge:
\[
ds^2 = a^2(\tau) \left[ -(1 + 2\psi) d\tau^2 + (1 - 2 \phi) \delta_{ij} dx^i dx^j \right],
\]
where $\tau$ is conformal time, $a(\tau) d\tau = dt$. A general tetrad consistent with this metric is written as~\cite{Golovnev:2018wbh}
\[
e^0_0 = a(\tau)(1 + \psi), \quad
e^0_i = a(\tau) \partial_i \xi, \quad
e^a_0 = a(\tau) \partial^a \xi, \quad
e^a_j = a(\tau) \left[ (1 - \phi) \delta^a_j + \epsilon_{ajk} \partial_k S \right],
\]
 where $\xi$ denotes the scalar part of the Lorentz boost, while $S$ corresponds to the scalar part of spatial rotations. As a result, two additional scalar variables appear beyond the metric perturbations. The variable $s$ is completely arbitrary and remains unconstrained, whereas $\xi$ is constrained and can be determined from the antisymmetric part of the field equations, which is non-trivial in this theory.

From the mixed component of the antisymmetric part of the field equations, we obtain in Fourier space
\begin{equation}\label{eq:zeta-Gslip}
k^2 \xi = 3\left(\phi' + \mathcal{H}\psi - \frac{\mathcal{H}' - \mathcal{H}^2}{\mathcal{H}} \phi \right).
\end{equation}

The remaining two equations arise from the diagonal components of the symmetric part of the field equations. The first one relates the two gravitational potentials,
\begin{equation} \label{per1}
\psi = \phi - 12\pi Q G \left(\frac{a}{k}\right)^2 (\rho + p)\sigma - \mathcal{H}\,\Xi\,\xi,
\end{equation}
where $Q = \frac{1}{f_T}$ and $\Xi = -12(\mathcal{H}' - \mathcal{H}^2) Q f_{TT}$. The final equation is obtained from the mixed component of the temporal--spatial part of the symmetric field equations and is given by
\begin{equation} \label{per2}
\phi' = 4\pi Q G \left(\frac{a}{k}\right)^2 (\rho + p)\theta
- \mathcal{H}\psi - \frac{\mathcal{H}\Xi}{a^2}\phi.
\end{equation}
The evolution of the tensor perturbations $h_{ij}$ is governed by
\begin{equation}
    h''_{ij}+\mathcal{H}\left(2+\frac{\Xi}{a^2}\right)h'_{ij}+k^2h_{ij}=0.
\end{equation}
Following the metric perturbations, the perturbed continuity and Euler equations for fluids are.
\begin{align}
\delta' &= -(1+w) (\theta - 3 \phi') + 3 \mathcal{H} \left( w - \frac{\delta p}{\delta \rho} \right) \delta, \\[2mm]
\theta' &= -\mathcal{H} (1 - 3 w) \theta - \frac{w'}{1 + w} \theta + \frac{\delta p / \delta \rho}{1 + w} k^2 \delta - k^2 \sigma + k^2 \psi.
\end{align}
These equations form the basis for calculating all relevant cosmological perturbation variables needed for data analysis. The background and perturbation equations in the framework of $f(T)$ gravity demonstrate that this theory constitutes a viable extension of the Teleparallel Equivalent of General Relativity (TEGR).

One of the main motivations for considering $f(T)$ gravity as an alternative to General Relativity (GR) and the standard $\Lambda$CDM cosmological model lies in the fact that it preserves the second-order nature of the field equations, unlike many other modified gravity theories which lead to higher-order equations. Consequently, $f(T)$ gravity naturally avoids the introduction of additional dynamical degrees of freedom in many cases, making it theoretically appealing as a minimal extension beyond GR within the cosmological framework. A wide range of cosmological models has been investigated within $f(T)$ gravity in order to describe the thermal and dynamical history of the Universe. In particular, several models have been proposed to account for the late-time acceleration without invoking a cosmological constant, see for example~\cite{Linder:2010py,Bengochea:2008gz,Nesseris:2013jea}. Another class of models is capable of generating an inflationary phase within the $f(T)$ framework~\cite{Awad:2017ign}. For more details on $f(T)$ theory and its applications see the reviews~\cite{Bahamonde:2021gfp,2013tgif.book.....A}.

%%%%%%%%%%%%%%%%%%%%%%%%%%%%%%%%%%%%%%%%%%%%%%%%%%%%%%%%%%%%%%%%%%%%%%%%%%%%%%%%%%%%%%%%%%%%%%%%%%%%%%%%%%%%%%%%%%%%%%%%%%%%%%%%%%%%%%
\section{New Exponential Infrared model}\label{models}
%%%%%%%%%%%%%%%%%%%%%%%%%%%%%%%%%%%%%%%%%%%%%%%%%%%%%%%%%%%%%%%%%%%%%%%%%%%%%%%%%%%%%%%%%%%%%%%%%%%%%%%%%%%%%%%%%%%%%%%%%%%%%%%%%%%%%%
In the general relativity context, the dark energy terminology has been introduced to explain the cosmic accelerated expansion of the late universe as confirmed by SNIa observations~\cite{Riess:1998AJ, Perlmutter:1999ApJ}. However, an adjustable weakening of gravity at late times on cosmic distances can mimic the same effect as dark energy. On the other hand, it recovers the successes of GR at early times and short distances at the solar system (milliparsec) or binary pulsar (microparsec) scales. This modification may be referred to as infrared corrections of GR gravity. In addition, these act effectively as a phantom DE without breaking the null energy condition~\citep{Carroll:2003st,Carroll:2004hc,Ludwick:2017tox}. It can also completely resolve the $H_0$ tension between the CMB and local measurements without violating the age constraints, even if tension remains with BAO measurements~\citep{El-Zant:2018bsc}.

Motivated by the phase portrait analysis of $f(T)$ teleparallel gravity, the exponential infrared $f(T)$ model has been suggested to produce viable cosmology~\cite{Awad:2017yod}
\begin{equation}\label{eq:expfT}
    f(T) = T  e^{\beta T_0/T},
\end{equation}
where $T_0=6H_0^2$ denotes the present value of the torsion scalar. Generally infrared corrections introduce new free parameters which may require further explanation and interpretation~\cite{Carroll:2006jn}. Remarkably, the above model describes a full spectrum of infrared corrections of gravity, while the model parameter $\beta$ turns out to be completely determined by the current values of the density parameters in terms of Lambert function. On another word, it does not introduce new free parameters, and therefore there is no specific value of $\beta$ can recover $\Lambda$CDM as a particular case. This is unlike other viable $f(T)$ theories~\cite{Nesseris:2013jea} which introduce new free parameters where $\Lambda$CDM can be recovered for a particular choice of the model parameter. This feature makes the exponential infrared $f(T)$ model~\eqref{eq:expfT} a genuine one. It has been shown that the model can produce two distinguishable phase space trajectories according to the chosen branch of the Lambert function~\cite{Awad:2017yod}. The principal branch model ($\beta>0$) has been confronted with observations~\cite{Hashim:2020sez,Hashim:2021pkq}. However, the secondary branch model ($\beta<0$) has been analyzed in Ref.~\citep{Akarsu:2024nas}. Both branches have been recently confronted with observational constraints~\cite{Hashim:2026yoy}.

In addition, an extension to this model, by introducing a power of the exponent term, has been considered~\cite{Santos:2022atq}. The extended model introduces an independent free parameter where the model can realize quintom dark energy behaviour at late times for some value of the new parameter. In Ref.~\cite{Santos:2022atq} the extended exponential infrared $f(T)$ gravity has been confronted with observational data on the background cosmic evolution. The full linear perturbation of the model with observational constraints from early and late universe~\cite{Hashim:2026M2}.

In the present work, we consider a particular case of the new exponential $f(T)$ teleparallel gravity~\cite{Hashim:2026M3}
\begin{equation}\label{eq:newexpfT}
    f(T) = \left(T +\alpha \sqrt{T T_0}\right) \exp\left(\beta \frac{T_0}{T}\right), \quad \text{where $\alpha=-1$}.
\end{equation}
In Ref.~\cite{Hashim:2026M3}, we show that the model parameter $\alpha\to -1$ when early + late observational constraints are used. For $\alpha\neq -1$, the model parameter $\beta$ is related to the torsional density parameter via Lambert-$W$ function where two branches of the solution are possible. Motivated by the results of Ref.~\cite{Hashim:2026M3} that $\alpha \to -1$, we assume the particular case $\alpha=-1$, which relates the parameter $\beta$ to the torsional density parameter in a simpler relation as we will shortly show. This nontrivially modifies~\eqref{eq:expfT} as we will show in the present study.

For the case $\alpha=-1$, it can be shown that the asymptotic behaviour of $f(T)$ in the limit ($T \to \infty$), which takes the form $f(T) \approx T - \sqrt{TT_0} + \beta T_0$. Given that the $\sqrt{TT_0}$ term serves as a total derivative in the background evolution, the model asymptotically recovers the dynamics of $\Lambda$CDM or $\Lambda_s$CDM (where $\Lambda_s<0$) in this limit, according to the sign of $\beta$. This consistency is also maintained at the perturbative level, as the functions $f_T$ and $f_{TT}$—which dictate the evolution of perturbations—exhibit the same $\Lambda$CDM-like asymptotic limits. Conversely, at late times, the exponential term becomes significant, which induces a cosmological behavior distinguishable from both $\Lambda$CDM and the models in~\cite{Awad:2017yod,Hashim:2020sez,Hashim:2021pkq,Hashim:2026yoy,Hashim:2026M2,Hashim:2026M3}.

Substituting the $f(T)$ function~\eqref{eq:newexpfT} into Friedmann equation~\eqref{fir_frid} after dividing by $3H_0^2$, we can express the normalized Hubble parameter as~\cite{Nesseris:2013jea}
\begin{equation}\label{eq:dimless_fTFR}
    E^2(z) = \Omega_{m0} (1+z)^3 + \Omega_{r0} (1+z)^4 + \Omega_{T0} \, y(E, \beta),
\end{equation}
where $E(z)=H(z)/H_0$, $\Omega_{m0}$ and $\Omega_{r0}$ denote the present values (i.e. $z=0$) of matter and radiation densities, while
\begin{equation}
    \Omega_{T0} = \frac{\kappa^2 \rho_T(0)}{3H_0^2}, \qquad
    y(E, \beta) = \frac{1}{T_0\Omega_{T0}}\left({T} + {f} - 2{T f_T}\right).
\end{equation}
For the $f(T)$ model under consideration~\eqref{eq:newexpfT}, we have
\begin{equation}
    y(E,\beta) = \frac{1}{\Omega_{T0}}\left[E^2 + \exp\left(\frac{\beta}{E^2}\right)
    \left( - E^2  + 2\beta - 2\frac{\beta}{E}\right)\right].
    \label{y_E}
\end{equation}
At the present epoch ($E=1$), we have the normalization condition $y(1)=1$,
which leads to a relation between $\beta$ and $\Omega_{T0}$:
\begin{equation}\label{eq:beta}
    \beta=ln(1-\Omega_{T0}).
\end{equation}
From the field equation at $z=0$, it follows that
\begin{equation}
    1 - \Omega_{T0} = \Omega_{m0} + \Omega_{r0}.
\end{equation}
Since $0<\Omega_{T0}<1$, we obtain that $\beta<0$ in the present model. Remarkably, the model parameter $\beta$ can be entirely expressed in terms of the density parameters. For example, if $\Omega_{T0}\sim 0.7$, we obtain $\beta\sim -1.2$ as recognized via Eq.~\eqref{eq:beta}. In this sense, the model does not introduce extra parameters into the parameter space, therefore we still have six parameters as in $\Lambda$CDM model to analyze the present $f$CDM model.

We note that the model parameter $\beta$ is always negative. This distinguishes the present model from the previous study~\cite{Hashim:2020sez} which considered the positive $\beta$ case of the principal branch of Lambert function solution. Indeed, the secondary branch solution (i.e. $\beta<0$) has been considered in a recent study~\cite{Akarsu:2024nas} which showed that the secondary branch solution realizes a smooth transition from negative to positive density. However, it overestimates $H_0$ value and cannot be considered as reliable model (see also~\cite{Hashim:2026yoy}). The same study shows that a consistent $H_0$ value can be produced by introducing a cosmological constant. In the present model, we show that the new exponential $f(T)$ gravity~\eqref{eq:newexpfT} can produce a smooth transition from negative to positive density of the dark energy with consistent $H_0$ value with no need to a cosmological constant.

%%%%%%%%%%%%%%%%%%%%%%%%%%%%%%%%%%%%%%%%%%%%%%%%%%%%%%%%%%%%%%%%%%%%%%%%%%%%%%%%%%%%%%%%%%%%%%%%%%%%%%%%%%%%%%%%%%%%%%%%%%%%%%%%%%%%%%
\section{Phase Portrait}\label{phase}
%%%%%%%%%%%%%%%%%%%%%%%%%%%%%%%%%%%%%%%%%%%%%%%%%%%%%%%%%%%%%%%%%%%%%%%%%%%%%%%%%%%%%%%%%%%%%%%%%%%%%%%%%%%%%%%%%%%%%%%%%%%%%%%%%%%%%%
By considering the relation~\eqref{eq:Tsc}, it has been shown that governing dynamical equation of $f(T)$ teleparallel cosmology can be written as a one dimensional autonomous system, for more details see~\cite{Awad:2017yod}
\begin{equation}
    \dot{H}    =    3(1+\omega)\,    \frac{f - H f_H}{f_{HH}},
\end{equation}
where $f_H=\frac{df(6H^2)}{dH}$ and $f_{HH}=\frac{d^2f(6H^2)}{dH^2}$. Substituting the $f(T)$ function~\eqref{eq:newexpfT} into the above equation, we obtain the corresponding dynamical system,
\begin{equation}\label{eq:Hdot}
\dot{H}= -\frac{3(1+\omega)}{2}\,\frac{H^4\!\left(H^3 - 2\beta H H_0^2 + 2\beta H_0^3\right)}{H^5 - \beta H_0^2 H^3 - \beta H_0^3 H^2 + 2\beta^{\,2} H_0^4 H - 2\beta^{\,2} H_0^5}.
\end{equation}
\begin{figure}
    \centering
    \includegraphics[width=0.7\linewidth]{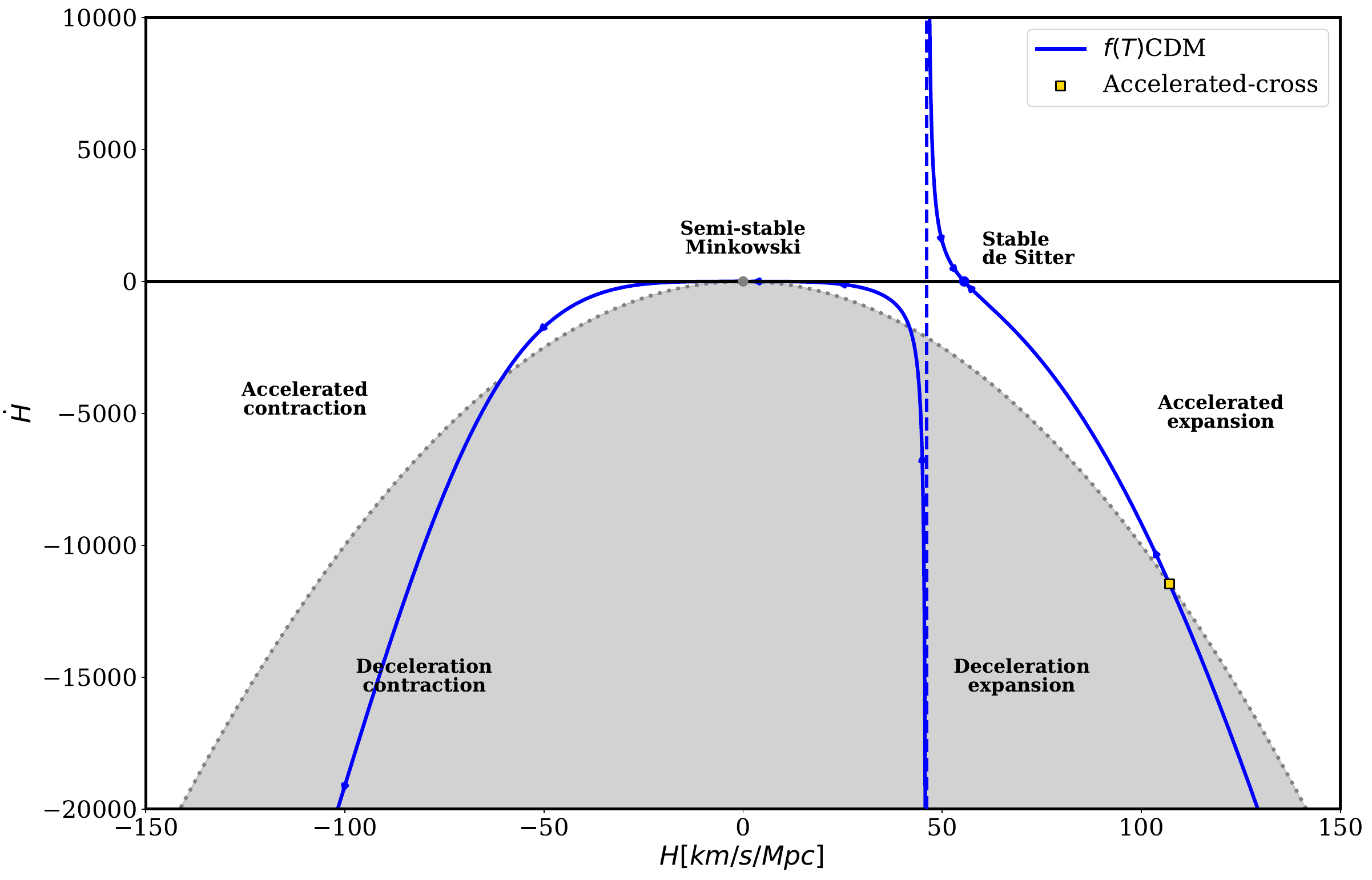}
    \caption{Phase portrait of the new exponential $f(T)$ model~\eqref{eq:newexpfT}. The dotted parabola represents the zero acceleration curve, $\dot{H}=-H^2$, where the shaded region represents the decelerated expansion (contraction) according to the sign of Hubble. At $H>0$ region, the trajectory begins with a finite time singularity (Big Bang) where $H\to \infty$ and $\dot{H}\to -\infty$, then it evolves toward a de Sitter fixed point $H^\ast$ (blue circle) crossing from decelerated to accelerated expansion (yellow square). We take $\Omega_{T0}=0.7$ and $H_0=70~{\rm km}\,{\rm s}^{-1}\,{\rm Mpc}^{-1}$.}
    \label{figphase}
\end{figure}
It can be shown that the leading term of the asymptotic expansion, at $H\to \infty$, requires
\begin{equation}
    \dot{H}= -\frac{3(1+\omega)}{2} H^2,
\end{equation}
which produces the GR phase portrait. This confirms that the present $f(T)$ model produces the GR limit at $H\to \infty$.

We note that the system admits four fixed points:
\begin{equation}
H = 0, \qquad H = H_0 x_i,
\end{equation}
where $x_i$ are the three roots of the cubic equation
\begin{equation}
x^3 - 2\beta x + 2\beta = 0.
\end{equation}
The number of solutions depends on the sign of $8\beta - 27$. If $8\beta - 27 > 0$, the system admits three de Sitter fixed points; otherwise, only one exists.
For $0 < \Omega_{T0} < 1$ we have $-\infty < \beta < 0$, and therefore the equation admits only one de Sitter fixed point. Figure~\ref{figphase} illustrates the phase portrait of the $f(T)$ model for $H_0 = 70~{\rm km}\,{\rm s}^{-1}\,{\rm Mpc}^{-1}$ and $\Omega_{T0} = 0.7$.

At the region $H < 0$, neither a fixed point nor singularity (divergence) in $\dot{H}$ are obtained. Since $\dot{H} \propto -H^2$ in the asymptotic limit $H \to -\infty$, the cosmic time is given by
\[
t-t_0 =- \int_{-\infty}^{H_0} \frac{dH}{\dot{H}},
\]
will be finite. Therefore, any trajectory begins at the $H < 0$ region, it evolves toward a big crunch singularity in a finite time.

At the region $H > 0$, there exists a de Sitter fixed point $H = H^\ast$ (represented by blue shaded circle in Figure~\ref{figphase}) and a sudden singularity (finite time singularity of Type II) at $H = H_{\rm s}$ where $\dot{H}$ diverges (represented by vertical dashed line in Figure~\ref{figphase}). The value of $H$ at which the sudden singularity occurs can be directly obtained from Eq.~\eqref{eq:Hdot} by requiring the vanishing of the denominator of its right hand side. For $H > H^\ast$, any trajectory starting at $H\to \infty$, i.e. with a big bang initial finite time singularity, evolves toward the fixed point $H^\ast$ in an infinite time where $\dot{H} < 0$ in this region.
For $H_{\rm  s} < H < H^\ast$, the system evolves toward the same fixed point $H^\ast$ where the universe is in a phantom region effectively (i.e., $\dot{H} > 0$), making $H^\ast$ a stable attractor. For $0 < H < H_{\rm s}$ the system approaches the Minkowski fixed point $H=0$ in infinite time, which makes Minkowski semi stable.
Although a singularity exists in this interval, the universe does not encounter it along its dynamical history. This is because, if the universe begins within that interval, it will be attracted to either the de Sitter fixed point or the Minkowski one, as we analyzed above.

The analysis is general because changing $H_0$ and $\Omega_{T0}$ merely shifts the fixed point locations without altering the qualitative behavior of the system.

%%%%%%%%%%%%%%%%%%%%%%%%%%%%%%%%%%%%%%%%%%%%%%%%%%%%
\subsection{Redshift of the deceleration acceleration transition}\label{redshift}
%%%%%%%%%%%%%%%%%%%%%%%%%%%%%%%%%%%%%%%%%%%%%%%%%%%%
We determine the zero acceleration curve, see the dotted parabola in Fig. \ref{figphase}, using the following constraint
\begin{equation}
\dot{H} = -H_{\rm acc}^2,
\end{equation}
where $H_{\rm acc}$ denotes the Hubble value at the decelerated to accelerated expansion. Substituting into the dynamical equation~\eqref{eq:Hdot}, it gives
\begin{equation}
-E_{acc}^2= -\frac{3(1+\omega)}{2}\,
\frac{
E_{acc}^4\!\left(E_{acc}^3 - 2\beta E_{acc} + 2\beta \right)
}{
E_{acc}^5 - \beta  E_{acc}^3 - \beta  E_{acc}^2
  + 2\beta^{\,2}  E_{acc} - 2\beta^{\,2}
},
\end{equation}
where $E_{\rm acc}=\frac{H_{\rm acc}}{H_0}$. At the matter dominated era, we have
\begin{equation} \label{beta}
    \beta=-\frac{(E_{\rm acc}-2+\sqrt{2E_{\rm acc}^2-5E_{\rm acc}+4})E_{\rm acc}^2}{2(E_{\rm acc}-1)}.
\end{equation}
Comparing the above constraint with~\eqref{eq:beta}, we obtain $E_{\rm acc}$ for a given value of $\Omega_{m0}$. The corresponding redshift $z_{\rm acc}$ can be obtained by solving Friedmann equation~\eqref{eq:dimless_fTFR}. This directly gives
\begin{equation}
z_{\rm acc}
=
\left(
\frac{E_{\rm acc}^2 - (1-\Omega_{m0})\, y(E_{\rm acc})}{\Omega_{m0}}
\right)^{1/3}
- 1.
\end{equation}
In Table \ref{tab:Hacc}, we obtain some numerical estimations of $z_{\rm acc}$ and $H_{\rm acc}$ for different choices of $\Omega_{m0}$ and $H_0$ applying Planck constraint $\Omega_{m0} h^2=0.142$.
\begin{table}[h!]
\centering
\begin{tabular}{c c c}
\hline
 $(\Omega_{m0},H_0)$ &  $z_{\rm acc}$ & $H_{\rm acc}$  \\
\hline
 (0.316,67) & 0.801 & 101.184  \\[3pt]
 (0.290,70) & 0.880 & 107.930  \\[3pt]
 (0.267,73) & 0.959 & 114.697  \\
\hline
\end{tabular}
\caption{The redshift and the corresponding Hubble values $(z_{\rm acc}, H_{\rm acc})$ at deceleration to acceleration expansion (dark energy dominated phase) for different $\Omega_{m0}$ and $H_0$ values. We fix the degeneracy relation $\Omega_{m0} h^2=0.142$.}
\label{tab:Hacc}
\end{table}

%%%%%%%%%%%%%%%%%%%%%%%%%%%%%%%%%%%%%%%%%%%%%%%%%%%%
\subsection{Phantom crossing and negative density}\label{phan}
%%%%%%%%%%%%%%%%%%%%%%%%%%%%%%%%%%%%%%%%%%%%%%%%%%%%

\begin{figure}[t!]
     \centering
     \begin{subfigure}[b]{0.48\textwidth}
         \centering
         \includegraphics[width=\textwidth]{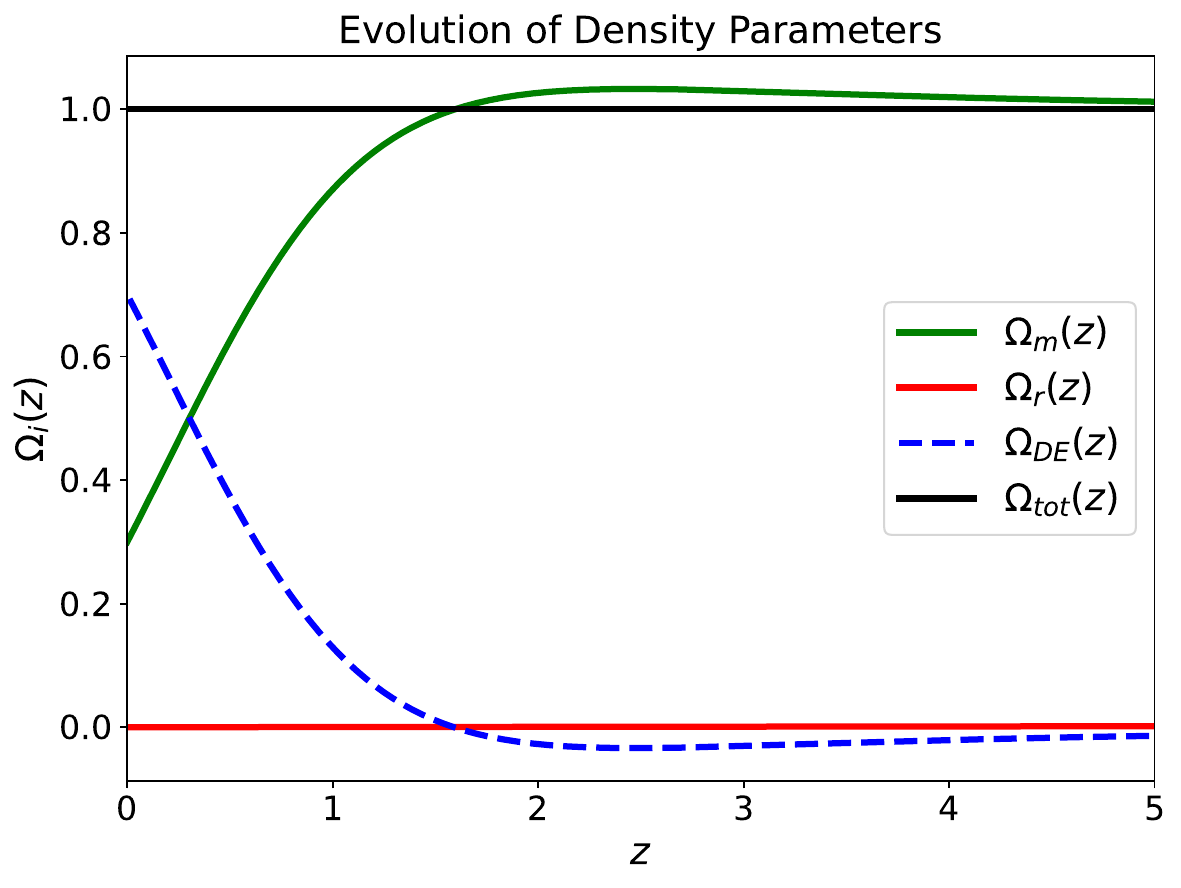}
         \caption{Evolution of density parameters $\Omega_i(z)$.}
         \label{fig:density_evolution}
     \end{subfigure}
     \hfill
     \begin{subfigure}[b]{0.48\textwidth}
         \centering
         \includegraphics[width=\textwidth]{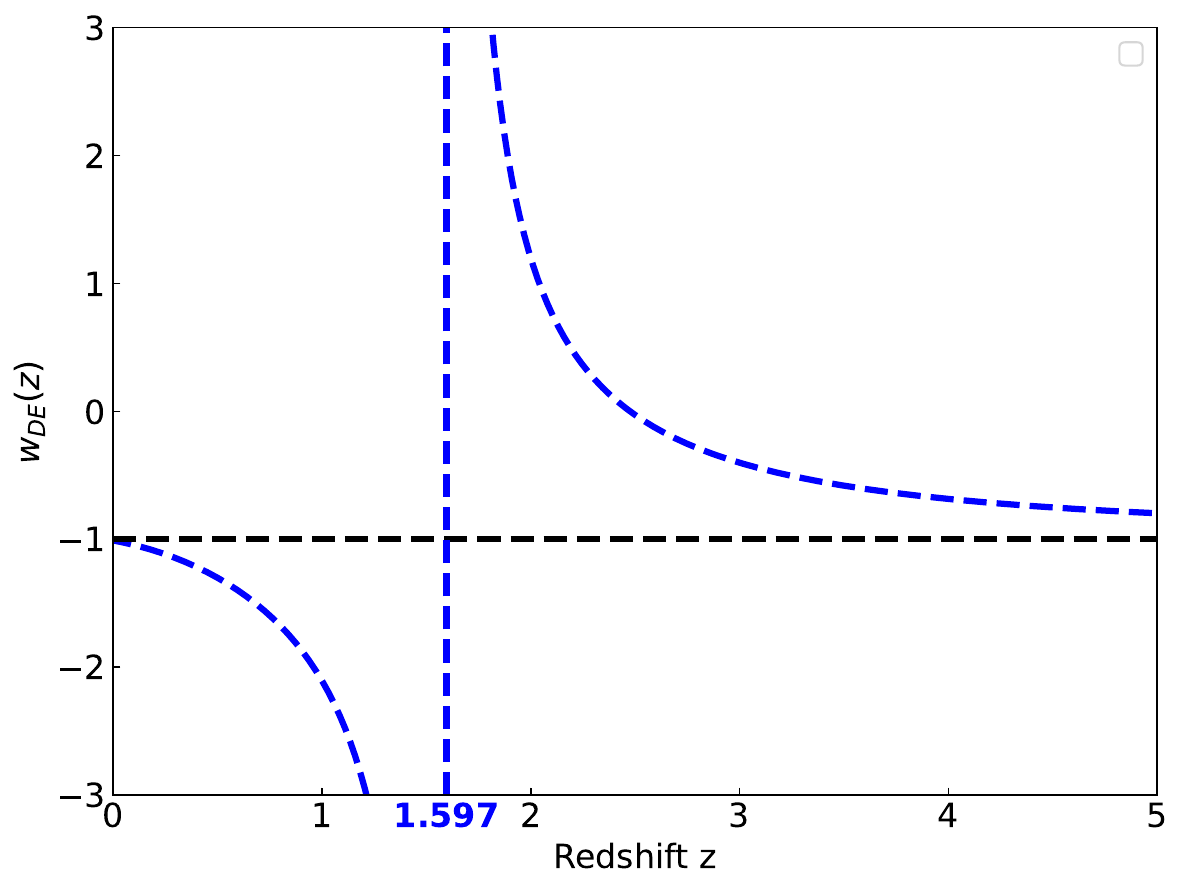}
         \caption{Evolution of the torsional EoS Parameter.}
         \label{fig:DEos}
     \end{subfigure}
     \caption{Cosmological evolution ($\Omega_{T0}=0.7$, $H_0=70~{\rm km}\,{\rm s}^{-1}\,{\rm Mpc}^{-1}$): (\subref{fig:density_evolution}) The density parameters~\eqref{eq:m_dens_para} and~\eqref{eq:T_dens_para} where $\Omega_{m}(z)$ exceeds unity at $z>1.597$ while the torsional dark energy density becomes negative. (\subref{fig:DEos}) The torsional EoS parameter~\eqref{eq:T_EoS} where a transition from non-phantom to phantom regime occurs at $z\sim 1.597$. At transition the torsion (DE) experiences a pole in its EoS (the torsion EoS diverges) as indicated by Eq.~\eqref{eq:div_EoS}. We note that any pathologies since $w_{DE}(z)$ is an effective EoS.}
     \label{fig:combined_results}
\end{figure}

Substituting the $f(T)$ function~\eqref{eq:newexpfT} into Eq.~\eqref{eq:newexp_tor}, and by using Eqs.~\eqref{eq:Tsc} and~\eqref{eq:newexp_Fr_eqns}, we write the matter density and pressure
\begin{eqnarray}
    \rho_m(H)&=&\frac{3H^2}{\kappa^2}\left(1-2\frac{\beta}{E^2}+2\frac{\beta}{E^3}\right)e^{\frac{\beta}{E^2}},\label{eq:m_dens}\\
     p_m(H)&=&-\frac{2\dot{H}}{\kappa^2}\left[1-\frac{\beta}{E^2}-\frac{\beta}{E^3}+2\frac{\beta^2}{E^4}-2\frac{\beta^2}{E^5}\right]e^{ \frac{\beta}{E^2}}-\rho_m(H).\label{eq:m_press}\qquad
\end{eqnarray}
Similarly, we give the torsional dark energy density and pressure
\begin{eqnarray}
    \rho_T(H)&=&\frac{3H^2}{\kappa^2}\left[1-\left(1-2\frac{\beta}{E^2}+2\frac{\beta}{E^3}\right)e^{\frac{\beta}{E^2}}\right],\label{eq:T_dens}\\
    p_T(H)&=&-\frac{3\beta H^2}{\kappa^2}\left[\frac{E^3-3E^2 +2\beta E-2\beta}{E^5-\beta E^3-\beta E^2 +2 \beta^2 E-2\beta^2}\right].\label{eq:T_press}\qquad
\end{eqnarray}
In the above we used the phase portrait trajectory~\eqref{eq:Hdot} at matter dominated era. Clearly, by substituting $\beta=0$ into the above equations $\rho_T$ and $p_T$ vanish and the GR case is restored. It follows by the torsional EoS parameter
\begin{equation}\label{eq:T_EoS}
      \omega_T=\frac{p_T}{\rho_T}=\frac{-\beta (E^3-3 E^2 +2\beta  E-2\beta )}{\left(E^5-\beta  E^3-\beta  E^2 +2 \beta^2  E-2\beta^2  \right) \left[1-\left(1- \frac{2\beta}{E^2}+\frac{2\beta}{E^3}\right)e^{ \frac{\beta}{E^2}}\right]}.
\end{equation}
We obtain the density parameters $\Omega_i(H)=\rho_i/\rho_c$, where the critical density $\rho_c=3H^2/\kappa^2$, as follows
\begin{eqnarray}
    \Omega_m(H)&=&\left[1-2\beta \left(\frac{1}{E}\right)^2+2\beta \left(\frac{1}{E}\right)^3\right]e^{ \frac{\beta}{E^2}},\label{eq:m_dens_para}\\
    \Omega_T(H)&=&1-\left[1-2\beta \left(\frac{1}{E}\right)^2+2\beta \left(\frac{1}{E}\right)^3\right]e^{ \frac{\beta}{E^2}}.\label{eq:T_dens_para}\qquad
\end{eqnarray}
The integration of the continuity equation~\eqref{eq:cont} at matter dominated era can be written as $\rho_m(H)=\rho_{m,0}/a(H)^3$ where $\rho_{m,0}$ is a constant of integration denoting the present value of the matter density. Thus, using Eq.~\eqref{eq:m_dens}, the scale factor reads
\begin{equation}
      a(H)=\left(\frac{\Omega_{m,0} H_0^2}{\Omega_m H^2}\right)^{1/3}=\frac{\Omega_{m,0}^{1/3} e^{-\frac{\beta }{3E^2}}}{\left(E^2-2\beta  + \frac{2\beta}{E}\right)^{\frac{1}{3}}},
\end{equation}
The redshift, using the relation $a(z)=\frac{1}{1+z}$, can be expressed in terms of $H$ as follows
\begin{equation}
    z(H)=\left(\frac{E^2-2\beta  +\frac{2\beta}{E}}{\Omega_{m,0} }\right)^{1/3}e^{\frac{\beta }{3E^2}}-1
\end{equation}
Using the inverse relation of the above equation, i.e. $H(z)$, one can evaluate the density parameters~\eqref{eq:m_dens_para} and~\eqref{eq:T_dens_para}. In Fig. \ref{fig:combined_results}(\subref{fig:density_evolution}), we plot the evolution of the density parameters. Similarly, we plot the evolution of the torsional dark energy EoS parameter~\eqref{eq:T_EoS} at different redshifts as shown in Fig. \ref{fig:combined_results}(\subref{fig:DEos}).

From the asymptotic expansion of $y(E)$, see Eq.~\eqref{y_E}, we find
\[
y(E)\to \frac{\beta}{\Omega_{T0}}
\quad\text{as } E\to\infty.
\]
Thus, for negative $\beta$, the dark energy density becomes negative at high redshift. Since $y(E)\to 1$ at present as $E\to 1$, the dark energy component must cross zero at some redshift as seen in Fig. \ref{fig:combined_results}(\subref{fig:density_evolution}). It proves convenient to write Friedmann equation for a general dark energy EoS as follows
\begin{equation}
     E^2(z) = \Omega_{m0} (1+z)^3 + \Omega_{r0} (1+z)^4 + \Omega_{DE0} \exp{\left(\int_0^z \frac{3(1+\omega_{DE}(\bar{z}))}{1+\bar{z}}d\bar{z}\right)},
\end{equation}

The comparison of the above equation with Eq.~\eqref{eq:dimless_fTFR} gives the following relation
\begin{equation}\label{eq:div_EoS}
    w_{DE}(z)=-1+\frac{1}{3}(1+z)\frac{d\ln{y(z)}}{dz}.
\end{equation}

Therefore, the torsional dark energy EoS parameter must diverge at the transition from negative density to positive density, i.e. $y(z_{tr})=0$, and it will be phantom region for $z<z_{tr}$ as seen in Fig. \ref{fig:combined_results}(\subref{fig:DEos}).

The transition of the dark energy density from negative to positive values can be determined from the condition $y(E)=0$. The corresponding transition redshift $z_{tr}$ is obtained as
\begin{equation}\label{z_t}
    z_{tr}=\left(\frac{E^2 }{\Omega_{m0}}\right)^{1/3}- 1.
\end{equation}
For $\Omega_{T0}=0.7$, we obtain $z_{tr}\sim 1.59$. It can be seen that the transition redshift $z_{tr}$ is completely determined by the fundamental cosmological parameters. Unlike various negative transition dark energy models that introduce additional parameters to handle the transition from negative to positive~\cite{Soriano:2025gxd,Toda:2024ncp,Gomez-Valent:2024ejh}. On the other hand, the value of $z_{tr}$ is in agreement with the Baryon Oscillation Spectroscopic Survey (BOSS) collaboration~\cite{BOSS:2014hhw} to solve the 2.5$\sigma$ tension of Ly-$\alpha$ BAO measurement at $z_{eff}=2.34$ with Planck-$\Lambda$CDM, where the dark energy density $\rho_{DE}<0$ at $z>1.6$. It is noteworthy that this behavior arises entirely from the modification of the underlying geometric framework used to describe the gravitational sector, and consequently from the modification of the gravitational field equations themselves. It is not the result of introducing an unknown dark energy component that requires a negative kinetic term as in phantom model or problematic physical mechanism to produce negative dark energy density.

%%%%%%%%%%%%%%%%%%%%%%%%%%%%%%%%%%%%%%%%%%%%%%%%%%%%%%%%%%%%%%%%%%%%%%%%%%%%%%%%%%%%%%%%%%%%%%%%%%%%%%%%%%%%%%%%%%%%%%%%%%%%%%%%%%%%%%
\section{Data and priors}\label{data}
%%%%%%%%%%%%%%%%%%%%%%%%%%%%%%%%%%%%%%%%%%%%%%%%%%%%%%%%%%%%%%%%%%%%%%%%%%%%%%%%%%%%%%%%%%%%%%%%%%%%%%%%%%%%%%%%%%%%%%%%%%%%%%%%%%%%%%
The parameter space for our analysis includes the usual six parameters of the standard $\Lambda$CDM model,  Table~\ref{tab:priors} shows the priors used for each parameter. To calculate the cosmological quantities , we use a modified version of the \texttt{CLASS} code~\cite{2011JCAP...07..034B} to consider $f(T)$ teleparallel gravity\protect\footnotemark[2].
\footnotetext[2]{
This code is publicly available at \href{https://github.com/mwhashim/class_tmg}{\faGithub} \url{https://github.com/mwhashim/class_tmg},
}
The cosmological parameters are sampled using the \texttt{Cobaya} ~\cite{Torrado:2020dgo} via an Markov Chain Monte Carlo (MCMC) algorithm. We run the chains until the Gelman-Rubin convergence criterion reaches $R-1 < 0.02$ for all the parameters of Table~\ref{tab:priors}. Finally, we use the \texttt{GetDist} package~\cite{Lewis:2019xzd} to analyze the output chains and plot the parameter posteriors and contours.
\begin{table}[h!]
\centering
\caption{Cosmological parameters and their priors used in the analysis.}
\begin{tabular}{|c|c|}
\hline
\textbf{Parameter} & \textbf{Prior} \\
\hline
$\Omega_{b0}h^2$ & $ \mathcal{U}[0.005, 0.03]$ \\
$\Omega_{c0}h^2$ & $ \mathcal{U}[0.01, 0.2]$ \\
$100\theta_s$ & $ \mathcal{U}[0.5, 10]$ \\
$ln(10^{10}A_s)$ & $ \mathcal{U}[2, 4]$ \\
$n_s$ & $\mathcal{U}[0.7, 1.2]$ \\
$\tau_{reio}$ & $ \mathcal{U}[0.01, 0.1]$ \\
\hline
\end{tabular}
\label{tab:priors}
\end{table}

In this work, we use several observational datasets to constrain our cosmological model. Each dataset is described below, including the likelihoods used in our analysis.

%%%%%%%%%%%%%%%%%%%%%%%%%%%%%%%%%%%%%%%%%%%%%%%%%%%%
\subsection{Cosmic Microwave Background (CMB)}
%%%%%%%%%%%%%%%%%%%%%%%%%%%%%%%%%%%%%%%%%%%%%%%%%%%%
Planck data represent measurements of the temperature anisotropies and polarization of the CMB provided by the Planck Collaboration (2018)~\cite{Planck:2018vyg}. In our analysis, we use both the high-$\ell$ and low-$\ell$ likelihoods, as well as the lensing likelihood:

\begin{itemize}
    \item \textbf{High-$\ell$ TT/TE/EE}: \texttt{Planck\_high\_l\_TTTEEE} covering $30 < \ell < 2508$ for TT and $30 < \ell < 1996$ for TE and EE.
    \item \textbf{Low-$\ell$ TT}: \texttt{Planck\_lowl\_TT} for temperature in the range $2 < \ell < 29$.
    \item \textbf{Low-$\ell$ EE}: \texttt{Planck\_lowl\_EE} for polarization in the same multipole range.
    \item \textbf{Lensing}~\cite{Planck:2018lbu}: \texttt{Planck\_lensing}, providing measurements of the projected gravitational potential derived from CMB lensing, offering an independent constraint on the growth of structure.
\end{itemize}

%%%%%%%%%%%%%%%%%%%%%%%%%%%%%%%%%%%%%%%%%%%%%%%%%%%%
\subsection{Baryon Acoustic Oscillations (BAO)}
%%%%%%%%%%%%%%%%%%%%%%%%%%%%%%%%%%%%%%%%%%%%%%%%%%%%

For BAO measurements, we use the data from the Dark Energy Spectroscopic Instrument (DESI) Year Two dataset~\cite{DESI:2025zgx}. In our analysis, we employ the corresponding likelihood \texttt{bao.desi\_dr2.desi\_bao\_all} in Cobaya. This dataset provides measurements of The Hubble parameter,The co-moving angular diameter distance and The volume-averaged distance

%%%%%%%%%%%%%%%%%%%%%%%%%%%%%%%%%%%%%%%%%%%%%%%%%%%%
\subsection{Type Ia Supernovae (SN~Ia)}
%%%%%%%%%%%%%%%%%%%%%%%%%%%%%%%%%%%%%%%%%%%%%%%%%%%%
Supernova Type Ia serve as standardizable candles through the measurement of the peak apparent magnitude $m_{B}$, which allows us to infer the luminosity distance and therefore the distance modulus. The supernova dataset provides the necessary light-curve information to compute the relative distance from $m_{B}$, which we use directly in our analysis. The datasets and their corresponding likelihoods in Cobaya are:

\begin{itemize}
    \item \textbf{Union3}~\cite{Rubin:2023jdq}: The latest release of the Union SN compilation series containing more than 2000 calibrated SN~Ia observations. \\
    \textit{Likelihood used:} \texttt{sn.Union3}
    \item \textbf{Pantheon+}~\cite{Brout:2022vxf}: An updated compilation containing approximately 1700 SN~Ia observations, making it one of the largest and most homogeneous SN datasets. Pantheon+ spans a wide redshift range ($0.01 < z < 2.3$). \\
    \textit{Likelihood used:} \texttt{sn.Pantheon\_Plus}
    \item \textbf{DES-SN}~\cite{DES:2024jxu}: The five-year Dark Energy Survey supernova sample, providing a high-quality set of approximately 300--400 SN~Ia. The dataset covers the redshift range $0.05 < z < 1.2$. \\
    \textit{Likelihood used:} \texttt{sn.desy5}
\end{itemize}

%%%%%%%%%%%%%%%%%%%%%%%%%%%%%%%%%%%%%%%%%%%%%%%%%%%%%%%%%%%%%%%%%%%%%%%%%%%%%%%%%%%%%%%%%%%%%%%%%%%%%%%%%%%%%%%%%%%%%%%%%%%%%%%%%%%%%%
\section{Results}\label{res}
\renewcommand{\arraystretch}{1.4}
%%%%%%%%%%%%%%%%%%%%%%%%%%%%%%%%%%%%%%%%%%%%%%%%%%%%%%%%%%%%%%%%%%%%%%%%%%%%%%%%%%%%%%%%%%%%%%%%%%%%%%%%%%%%%%%%%%%%%%%%%%%%%%%%%%%%%%
Tables~\ref{tab:LCDM} and~\ref{tab:f1} show the best-fit values and 68\% confidence intervals for the primary cosmological parameters listed in Table~\ref{tab:priors}, as well as the four derived parameters $H_0$, $\sigma_8$, $\Omega_{m0}$, and $S_8$ for both $\Lambda\text{CDM}$ and $f$CDM models, we also consider $z_{tr}$, which denotes the redshift value at which the dark energy density transitions from negative to positive in the $f$CDM model.
\begin{table}[H]
\centering
\caption{\small The 68\% confidence intervals of the bestfit cosmological parameters of the $\Lambda$CDM model for different combinations of datasets, mentioned in Sec. \ref{data}, with the corresponding minimum $\chi^2$ values.}
\resizebox{\textwidth}{!}{
\begin{tabular}{lcccccc}
\toprule
Parameter & Planck & Planck+DESI & Planck+DESI+lensing & Planck+DESI+lensing+Union3 & Planck+DESI+lensing+PantheonPlus & Planck+DESI+lensing+DES \\
\midrule
$\Omega_{b0}h^2$ & $0.02234\pm0.00015$ & $0.02252\pm0.00013$ & $0.02253\pm0.00012$ & $0.022250\pm0.00013$ & $0.02250\pm0.00013$ & $0.02249\pm0.00013$ \\[2pt]
$\Omega_{c0}h^2$  & $0.1202\pm0.0013$ & $0.11753\pm0.00067$ & $0.11767\pm0.00063$ & $0.1178\pm0.00065$ & $0.11789\pm0.00065$ & $0.11810\pm0.00064$ \\[2pt]
$100\theta_s$  & $1.04185\pm0.00030$ & $1.04209\pm0.00027$ & $1.04209^{+0.00029}_{-0.00026}$ & $1.04209\pm0.00028$ & $1.04208\pm0.00028$ & $1.04204\pm0.00028$\\[2pt]
$n_s$ & $0.9643\pm0.0043$ & $0.9707\pm0.0033$ & $0.9704\pm0.0032$ & $0.9697\pm0.0035$ & $0.9697\pm0.0033$ & $0.9694\pm0.0033$ \\[2pt]
$\ln(10^{10}A_s)$ & $3.044\pm0.016$ & $3.046\pm0.017$ & $3.052\pm0.015$ & $3.052\pm0.015$ & $3.052\pm0.015$ & $3.051\pm^{+0.014}_{-0.016}$ \\[2pt]
$\tau_{reio}$ & $0.0540\pm0.0078$ & $0.0578^{+0.0071}_{-0.0083}$ & $0.0606\pm0.0071$ & $0.0599\pm0.0073$& $0.0601\pm0.0074$ & $0.0595^{+0.0070}_{-0.0078}$ \\[2pt]
\midrule
$H_0~{\rm km}\,{\rm s}^{-1}\,{\rm Mpc}^{-1}$ & $67.28\pm0.6$ & $68.46\pm0.30$ & $68.42\pm0.29$ & $68.32\pm0.29$ & $68.31\pm0.30$ & $68.21\pm0.28$ \\[2pt]
$\Omega_{m0}$ & $0.3162\pm0.0086$ & $0.3002\pm0.0038$ & $0.3009\pm0.0036$ & $0.3022\pm0.0037$ & $0.3021\pm0.0037$ & $0.3035\pm0.0036$ \\[2pt]
$S_8$ & $0.833\pm0.016$ & $0.8049\pm0.0098$ & $0.8089\pm0.0086$ & $0.8110\pm0.0083 $& $0.8112\pm0.0084$ & $0.8132\pm0.0083$ \\[2pt]
$\sigma_8$ & $0.8113\pm0.0074$ & $0.8046^{+0.0066}_{-0.0074}$ & $0.8077\pm0.0062$ & $0.8081\pm0.0061$ & $0.8082\pm0.0061$ & $0.8084\pm0.0062$ \\[2pt]
\midrule
$\chi^2$ & $2808.649$ & $2825.2674$ & $2834.4092$ & $2862.9566$ & $4242.1026$ & $4484.5508$ \\
\bottomrule
\end{tabular}
\label{tab:LCDM}
}
\end{table}
\renewcommand{\arraystretch}{1.4}
\begin{table}[H]
\centering
\caption{
\small
The 68\% confidence intervals of the bestfit cosmological parameters of the $f$CDM model for different combinations of datasets, mentioned in Sec. \ref{data}, with the corresponding minimum $\chi^2$ values, where $\Delta \chi^2=\chi^2_f-\chi^2_\Lambda$.}
\resizebox{\textwidth}{!}{
\begin{tabular}{lcccccc}
\toprule
Parameter & Planck & Planck+DESI & Planck+DESI+lensing & Planck+DESI+lensing+Union3 & Planck+DESI+lensing+PantheonPlus & Planck+DESI+lensing+DES \\
\midrule
$\Omega_{b0}h^2$ & $0.02217\pm0.00014$ & $0.02179\pm0.00012$ & $0.02181\pm0.00012$ & $0.02179\pm0.00012$ & $0.02178\pm0.00012$ & $0.02176\pm0.00012$ \\[2pt]
$\Omega_{c0}h^2$  & $0.1212\pm0.0014$ & $0.12666\pm0.00066$ & $0.12684\pm0.00065$ & $0.12711\pm0.00064$ & $0.12719\pm0.00064$ & $0.12743\pm0.00063$ \\[2pt]
$100\theta_s$  & $1.04176\pm0.00030$ & $1.04134\pm0.00027$ & $1.04132\pm0.00027$ & $1.04130\pm0.00028$ & $1.04131\pm0.00028$ & $1.04130\pm0.00028$ \\[2pt]
$n_s$ & $0.9630\pm0.0045$ & $0.9504\pm0.0043$ & $0.9499\pm0.0032$ & $0.9491\pm0.0033$ & $0.9489\pm0.0033$ & $0.9485\pm0.0032$ \\[2pt]
$\ln(10^{10}A_s)$ & $3.049\pm0.016$ & $3.049\pm0.016$ & $3.061\pm0.014$ & $3.060\pm0.014$ & $3.065\pm0.014$ & $3.059\pm0.014$ \\[2pt]
$\tau_{reio}$ & $0.0555\pm0.0079$ & $0.0492^{+0.0077}_{-0.0070}$ & $0.0544\pm0.0070$ & $0.0537\pm 0.0060$ & $0.0533\pm0.0067$ & $0.0528\pm0.0067$ \\[2pt]
\midrule
$H_0~{\rm km}\,{\rm s}^{-1}\,{\rm Mpc}^{-1}$ & $74.48\pm0.61$ & $72.13\pm0.28$ & $72.08\pm0.27$ & $71.96\pm0.27$ & $71.93\pm0.28$ & $71.83\pm0.26$ \\[2pt]
$\Omega_{m0}$ & $0.2597\pm0.0065$ & $0.2866\pm0.0033$ & $0.2874\pm0.0032$ & $0.2888\pm0.0032$ & $0.2892\pm0.0032$ & $0.2904\pm0.0031$ \\[2pt]
$S_8$ & $0.787\pm0.015$ & $0.8444\pm0.0096$ & $0.8512\pm0.0063$ & $0.8537\pm0.0084$ & $0.8544\pm0.0084$ & $0.8567\pm0.0082$ \\[2pt]
$\sigma_8$ & $0.8460\pm0.0085$ & $0.8639\pm0.0074$ & $0.8698\pm0.0063$ & $0.8701\pm0.0063$ & $0.8702\pm0.0062$ & $0.8708\pm0.0061$ \\[2pt]
$z_{tr}$ & $1.744\pm0.027$ & $1.639\pm0.012$ & $1.636\pm0.012$ & $1.631\pm0.012$ & $1.629\pm0.011$ & $1.625\pm0.011$ \\[2pt]
\midrule
$\chi^2$ & $2845.3$ & $2914.92$ & $2933.24$ & $2976.16$ & $4357.74$ & $4612.54$ \\[2pt]
$\Delta\chi^2$ & $36.66$ & $89.66$ & $98.84$ & $113.21$ & $115.64$ & $127.99$ \\
\bottomrule
\end{tabular}
\label{tab:f1}
}
\end{table}

\begin{figure}%[h!]
    \centering
    \includegraphics[width=0.95\textwidth]{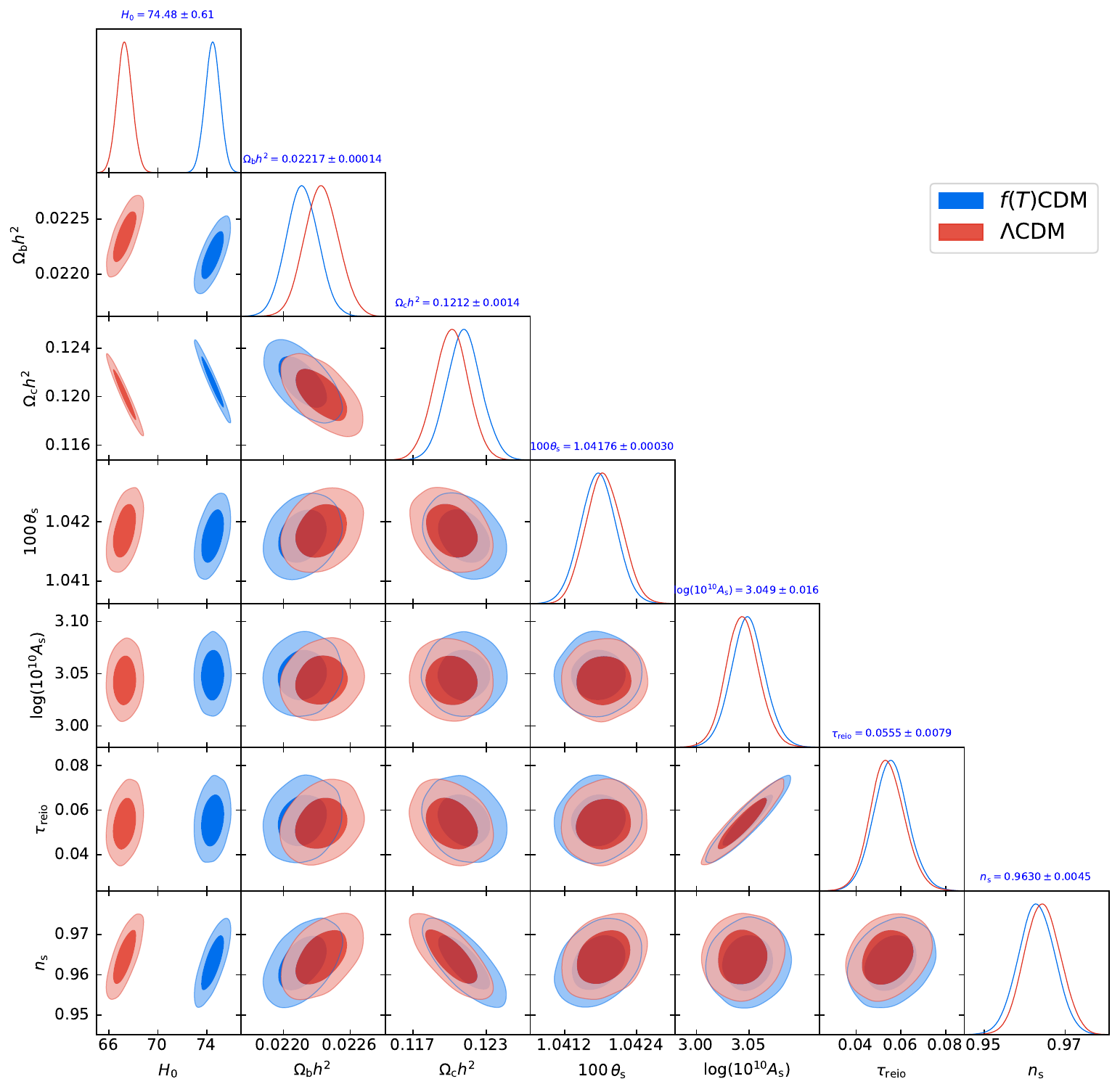}
    \caption{The 68\% (dark color) and 95\% (faint color) confidence level (CL) contours of the cosmological parameters for the $f$CDM vs $\Lambda$CDM models using Planck data set We also include the one-dimensional posterior distributions for the parameters.}
    \label{fig:f1_plk_lambda}
\end{figure}

\begin{figure}
    \centering
    \includegraphics[width=0.5\textwidth]{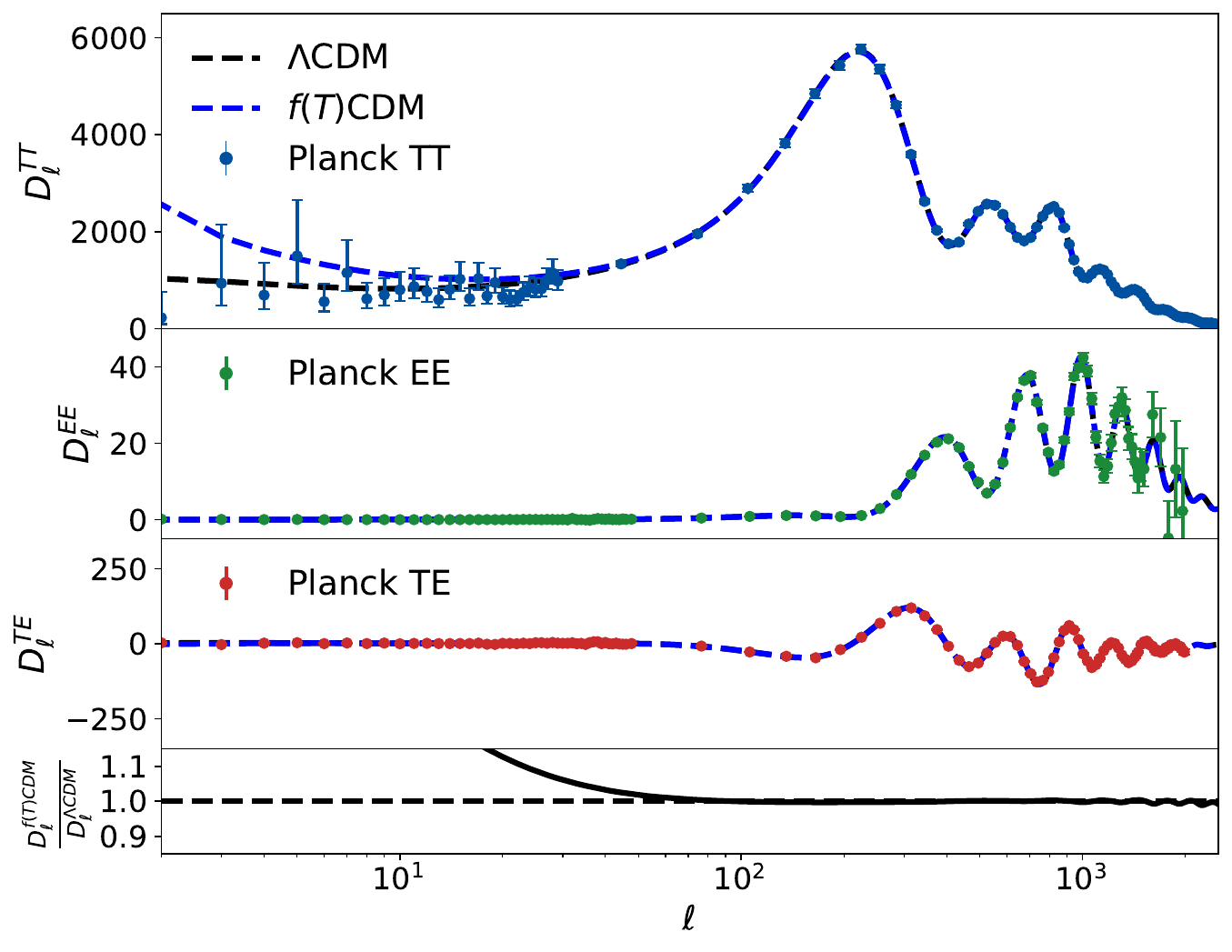}
    \caption{Temperature and polarization anisotropies, and their cross-correlations, for the $f$CDM and $\Lambda$CDM models. The solid and dashed lines represent the theoretical predictions constrained by \textit{Planck} data, while the points denote the actual \textit{Planck} measurements. The lower panel displays the ratio between the two models specifically for the temperature (TT) power spectrum.}
    \label{fig:placeholder}
\end{figure}

\begin{figure}[htbp]
    \centering
    \begin{subfigure}{0.4\linewidth}
      \centering
      \includegraphics[width=\linewidth]{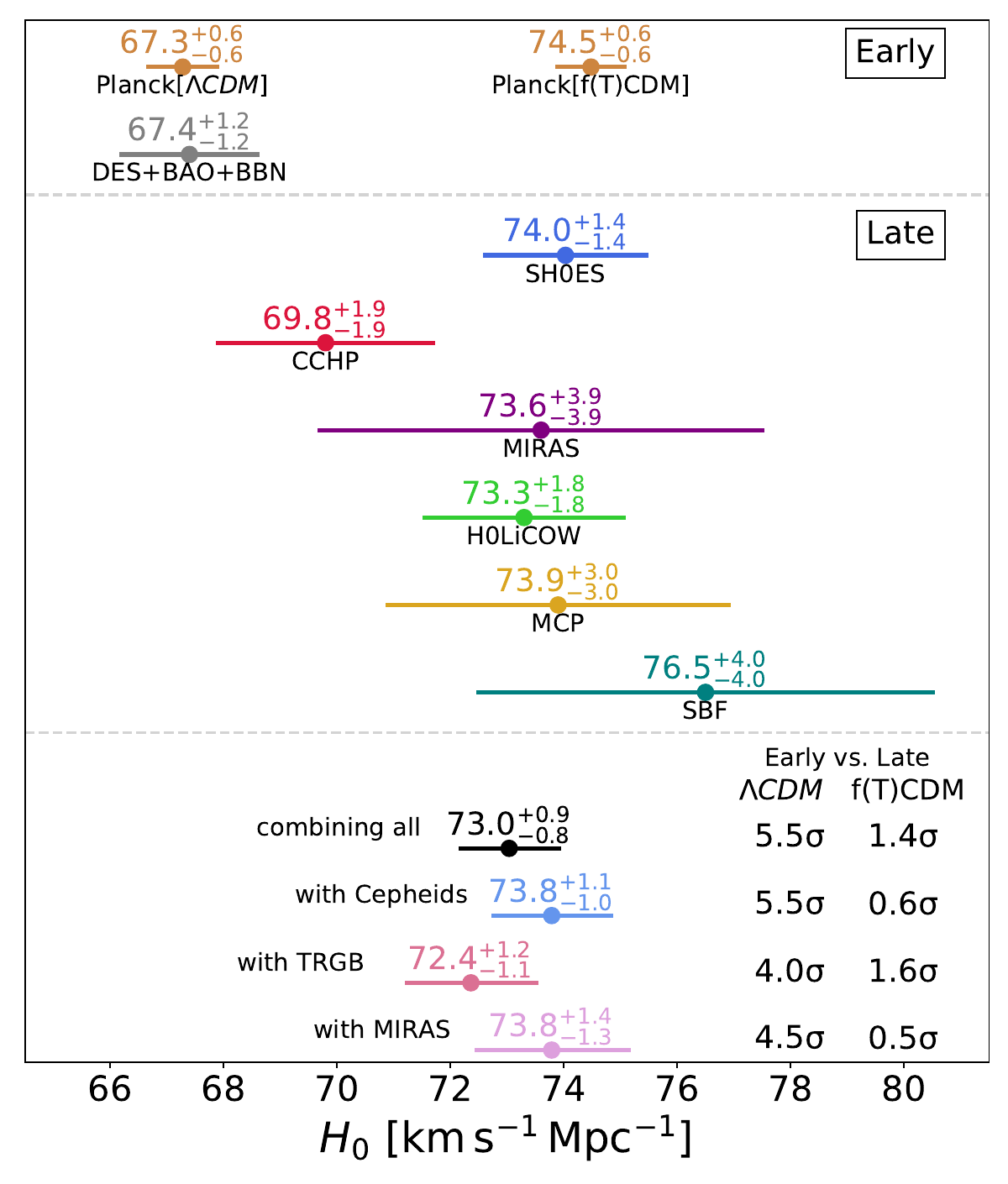}
      \caption{ $H_0$-tension}
      \label{fig:H0tens}
    \end{subfigure}\hspace{0.5cm}
    \begin{subfigure}{0.4\linewidth}
        \centering
        \includegraphics[width=\linewidth]{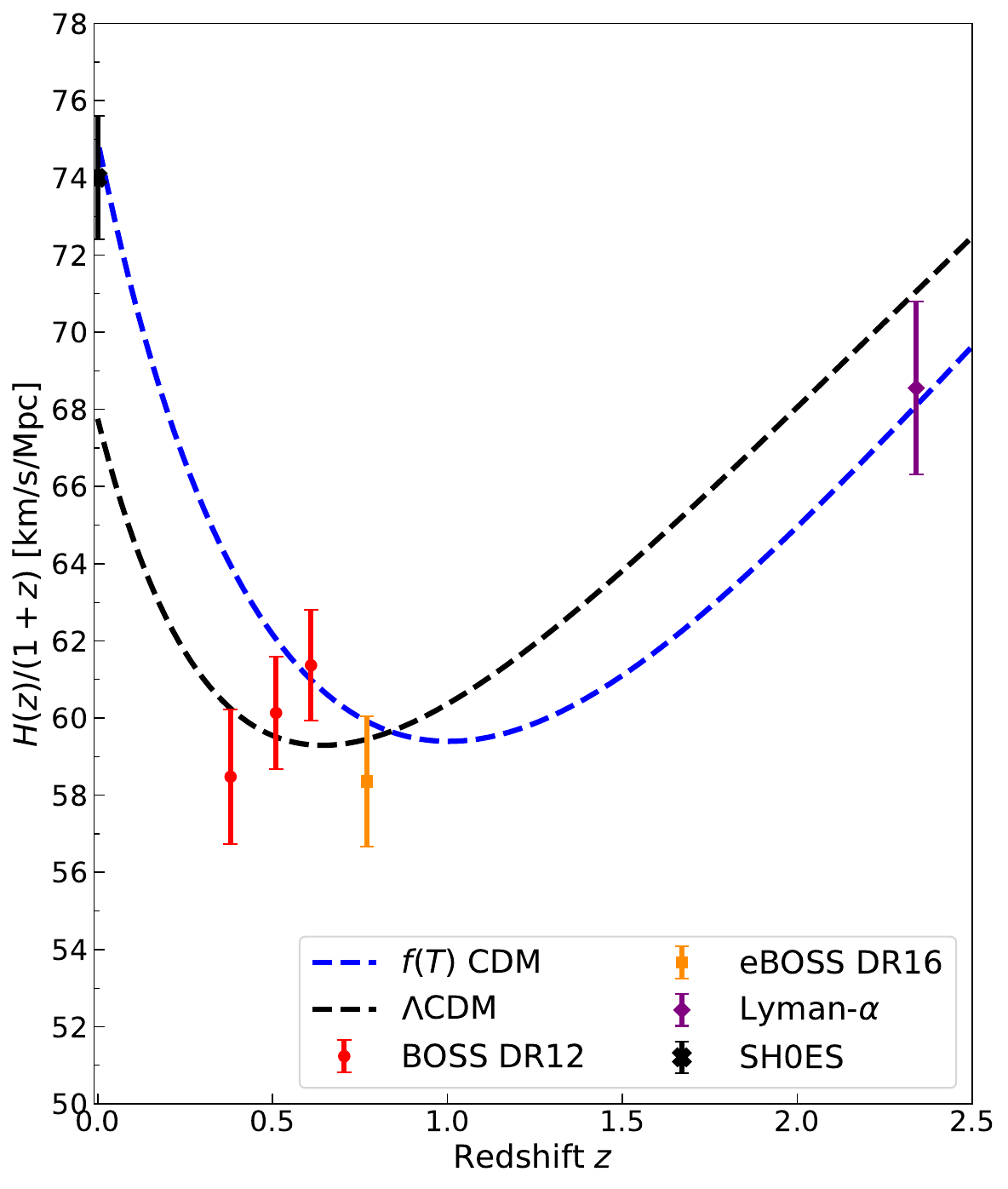}
        \caption{Hubble evolution}
        \label{fig:fTvLam}
    \end{subfigure}
    \caption{ (\subref{fig:H0tens})  Comparison of the $H_0$ tension between $f$CDM and $\Lambda\text{CDM}$ models\protect\footnotemark[3]. Constraints are inferred from \textit{Planck} 2018 data (CMB only) and various local measurements. The $f$CDM model significantly reduces the tension to $\approx 1\sigma$ by predicting a higher $H_0$ value, driven by the modified expansion history and the phantom-divide crossing in the late Universe. (\subref{fig:fTvLam}). We compare the evolution of $H(z)/(1+z)$ for the $\Lambda$CDM model (dashed black line) and the $f$CDM model (dashed blue line), using the best-fit values of cosmological parameters estimated from \textit{Planck} data, together with the BAO dataset from Boss DR12~\cite{BOSS:2016wmc}, eBoss DR16 ~\cite{eBOSS:2020xwt} and Ly-$\alpha$ from eBOSS DR14~\cite{eBOSS:2019ytm}.}
    \label{fig:hbblcmprs}
\end{figure}

For the Planck dataset, the two models exhibit almost identical best-fit values for the primary parameters within $<1\sigma$, see Fig. \ref{fig:f1_plk_lambda}.
This is due to the nearly identical behavior of the two models in the early Universe. We plot the CMB powerspectra (TT, TE and EE) for $\Lambda$CDM and the $f(T)$ model in Fig. \ref{fig:placeholder}. The plots show clearly the identical behavior of the models with a remarkable deviation of the $f(T)$ model from $\Lambda$CDM at low-$\ell$ ($\ell\lesssim 30$) in the TT power spectrum. This could be understood since these scales are related to the dark energy at large distances, where $f(T)$ model acts differently from $\Lambda$CDM. In particular, a large gravitational slip $\propto 1/k^2$ in the $f(T)$ model is expected at large scales as indicated by Eq.~\eqref{eq:zeta-Gslip}, see also~\cite{Li:2011wu}. However, for the derived parameters, the $f$CDM model predicts a higher value of $H_0 = 74.48 \pm 0.61~{\rm km}\,{\rm s}^{-1}\,{\rm Mpc}^{-1}$, which reduces the tension with local measurements~\cite{Riess:2016jrr} from $5\sigma$ in $\Lambda$CDM to roughly $1\sigma$ in $f$CDM. In Figure \ref{fig:hbblcmprs}(\subref{fig:H0tens}), we determine the $H_0$-tension level with different combinations of late universe data. Similarly, the cosmological parameter $S_8=\sigma_8 \sqrt{\Omega_{m0}/0.3}=0.787 \pm 0.015$ value in the $f$CDM model is lower than $\Lambda$CDM, $S_8=0.833 \pm 0.016$, when only Planck data without lensing data is used. This is due to the significant small $\Omega_{m0}$ in the $f$CDM model relative to $\Lambda$CDM. Otherwise, when late universe data is added, $S_8$ becomes larger in the $f(T)$ model in comparison to $\Lambda$CDM. However, it should be noted that $\sigma_8$ in the $f$CDM model is higher relative to $\Lambda$CDM in all data combinations. This will be discussed in Subsection \ref{sec:sigma8}.

\footnotetext[3]{Figure \ref{fig:hbblcmprs}(\subref{fig:H0tens}) is generated after modifying the code introduced by Vivien Bonvin and Martin Millon, available at \href{https://github.com/vbonvin/H0_tension} {\faGithub} \url{https://github.com/vbonvin/H0_tension}, in order to implement the model discussed in this work.}

\begin{figure}
    \centering
    \includegraphics[width=\textwidth]{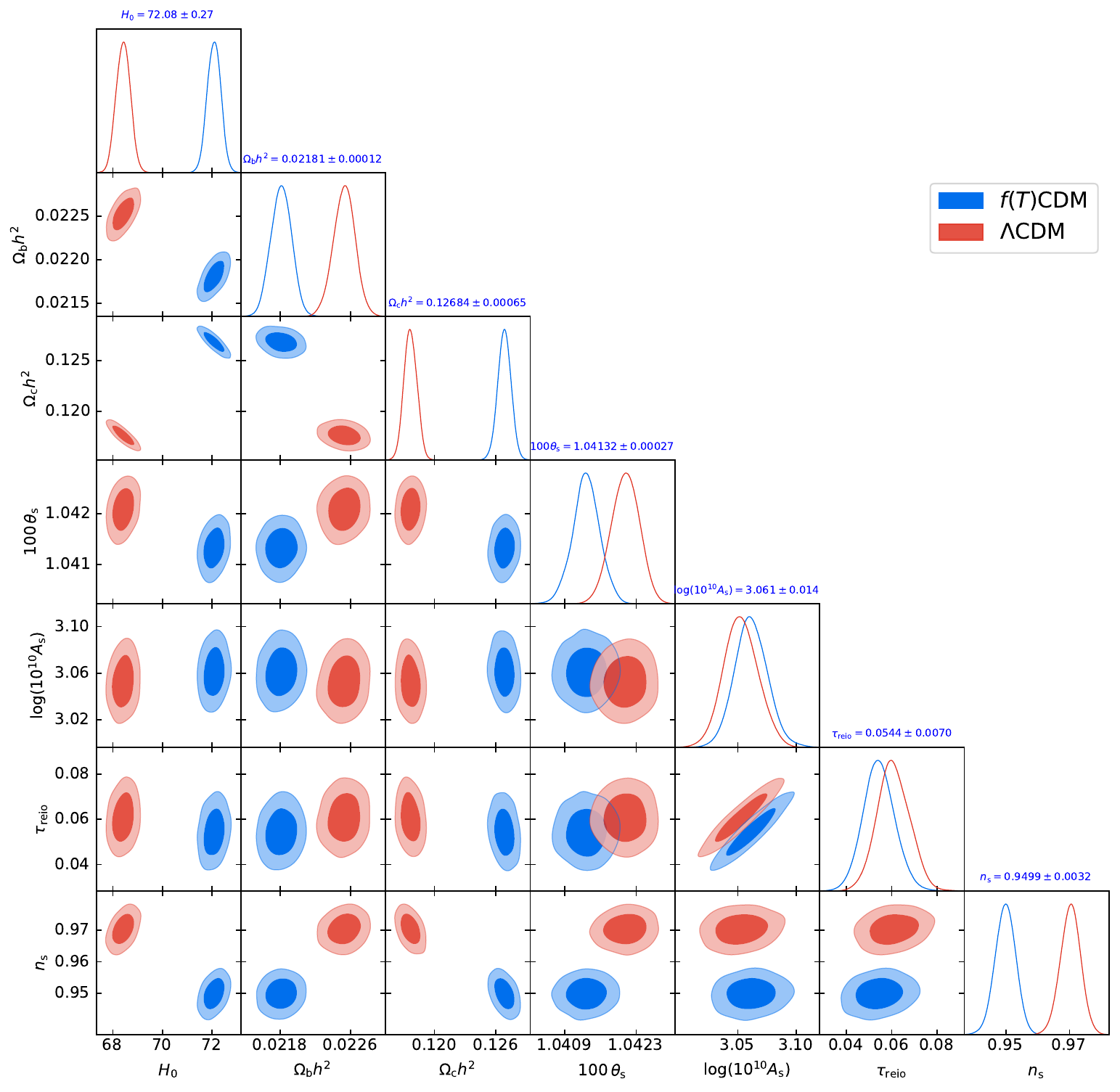}
    \caption{The 68\% (dark color) and 95\% (faint color) confidence level (CL) contours of the cosmological parameters for the $f$CDM vs $\Lambda $CDM models using Planck+DESI+lensing data set We also include the one-dimensional posterior distributions for the parameters.}
    \label{fig:f1_lambda}
\end{figure}

\begin{figure}
    \centering
    \includegraphics[width=\textwidth]{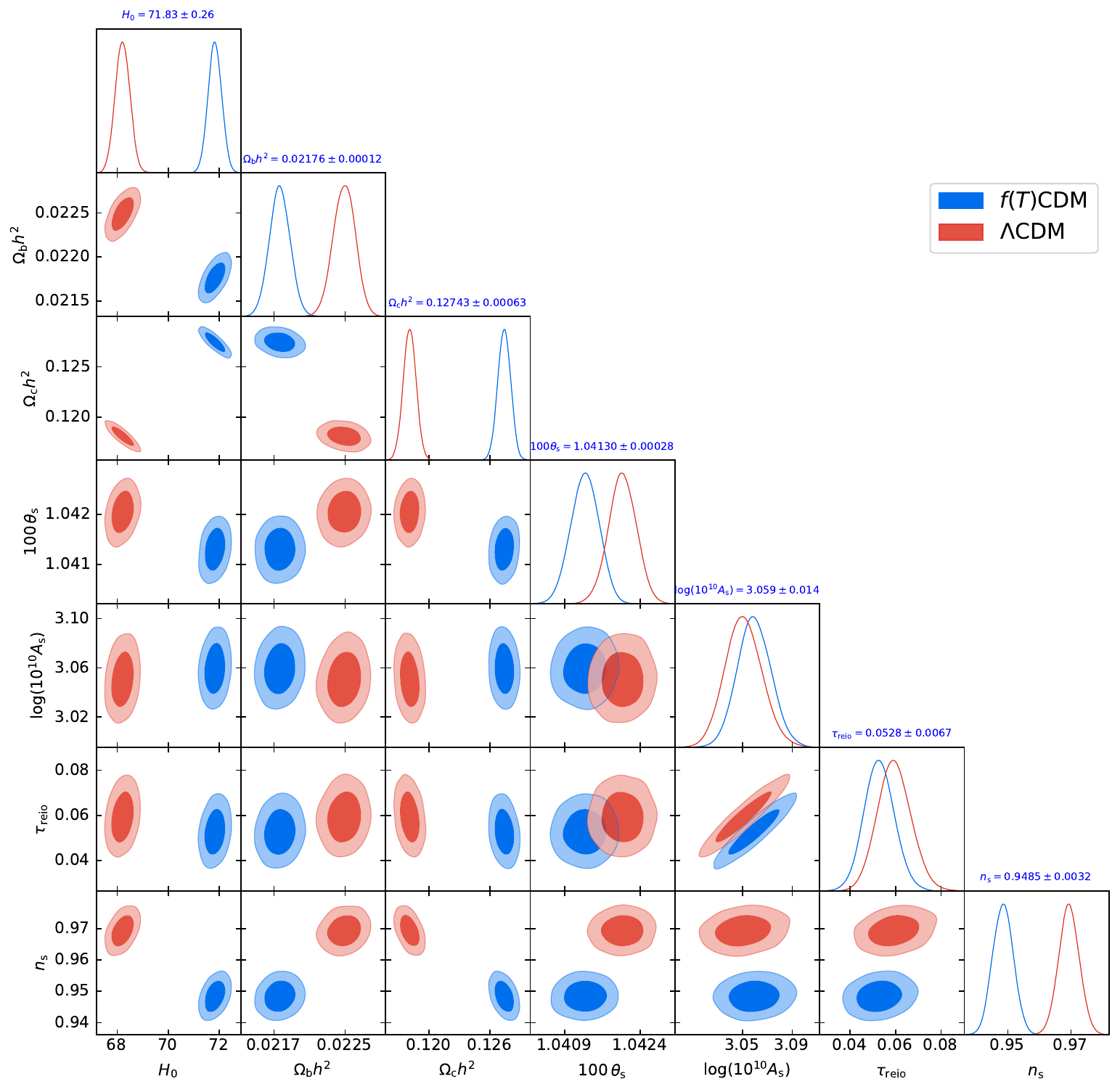}
    \caption{The 68\% (dark color) and 95\% (faint color) confidence level (CL) contours of the cosmological parameters for the $f$CDM vs $\Lambda $CDM models using Planck+DESI+lensing+DES data set We also include the one-dimensional posterior distributions for the parameters.}
    \label{fig:f1_DES_lambda}
\end{figure}

Additionally, a tension of $\sim 3.6 \sigma$ in $H_0$ persists for the $f$CDM model between Planck-only and Planck+DESI datasets, see Fig. \ref{fig:f1_lambda}. Adding supernova data (e.g., Union3/ PantheonPlus/ DES) largely reconciles the parameters, making them consistent with the combination of Planck+DESI, but same tension with Planck data alone, see Fig. \ref{fig:f1_DES_lambda}.

Notably, the transition redshift $z_{tr}$ largely depends on $\Omega_{m0}$, as can be seen from Eq.~\eqref{z_t}, where $z_{tr}$ decreases as $\Omega_{m0}$ increases when going from Planck-only to Planck+DESI data sets as seen in Table \ref{tab:f1}. For all data combinations we find that the torsional dark energy density is negative at $z_{tr}\gtrsim 1.6$. A key empirical motivation for presence of negative dark energy density arises from the BAO measurements by BOSS collaboration~\cite{BOSS:2014hhw}. In particular, the Ly-$\alpha$ observations show a level of tension at $\sim 2.5\sigma$ with $\Lambda$CDM, where a model independent reconstruction of the dark energy density provides a direct evidence that observational data requires $\rho_{DE}<0$ at $z_{tr}\gtrsim 1.6$, which evidently can be naturally reconciled within the present $f(T)$ model as seen in Table \ref{tab:f1}. In Fig. \ref{fig:hbblcmprs}(\subref{fig:fTvLam}), we plot the Hubble evolution of $\Lambda$CDM and $f$CDM according to the bestfit parameters as inferred using Planck data alone. The plots show the tensions between $\Lambda$CDM and BOSS Ly-$\alpha$ and SH0ES observations, whereas these tensions are absent within the present $f$CDM model.

Since both $\Lambda$CDM and the present $f(T)$ model have the same parameter space, the comparison between the two model can be performed via $\Delta\chi^2=\chi^2_{fCDM}-\chi^2_{\Lambda CDM}$ test. The results, in Table \ref{tab:f1}, show that the $f(T)$ model is strongly disfavored in comparison to $\Lambda$CDM in terms of the values of $\chi^2$ of the best fit parameters.

%%%%%%%%%%%%%%%%%%%%%%%%%%%%%%%%%%%%%%%%%%%%%%%%%%%%
\subsection{$n_s$-tension}\label{sec:nstension}
%%%%%%%%%%%%%%%%%%%%%%%%%%%%%%%%%%%%%%%%%%%%%%%%%%%%

We note that genuine tensions appear in the $f$CDM and $\Lambda$CDM models in some primary parameters, when late-time datasets such as DESI are combined with Planck data. In particular $\omega_{b0}=\Omega_{b0} h^2$, $\omega_{c0}=\Omega_{c0} h^2$, and $n_s$ as seen in Tables \ref{tab:LCDM} and \ref{tab:f1}. In the $f$CDM framework, there is a shift toward lower $n_s$ values, $\sim 2\sigma$, when DESI data are combined with Planck. A similar tension appears in the $\Lambda$CDM model $\sim 1.2\sigma$. However, in this case the shift occurs in the opposite direction towards higher $n_s$ value in agreement with~\cite{Ferreira:2025lrd}.

The apparent tension in the inflationary parameter $n_s$ by including DESI data is related to the degeneracy between $\omega_{m0}=\omega_{b0}+\omega_{c0}$ and $n_s$ in Planck measurements~\cite{Efstathiou:1998xx}. Planck data constrain two important combinations: one related to the location of the acoustic peaks, $\theta_s$, which is nearly model-independent, and another related to the height of the peaks, approximately $\Omega_{m0} h^2$, which exhibits degeneracy with $n_s$, particularly affecting the height of the first peak. On the other hand, DESI data provides strong constraints on $(\Omega_{m0}, r_s h)$. If DESI favors a larger value of $\Omega_{m0}$ compared to Planck alone, this directly affects the parameter combinations constrained by Planck.

\begin{figure}
    \centering
    \includegraphics[width=1\linewidth]{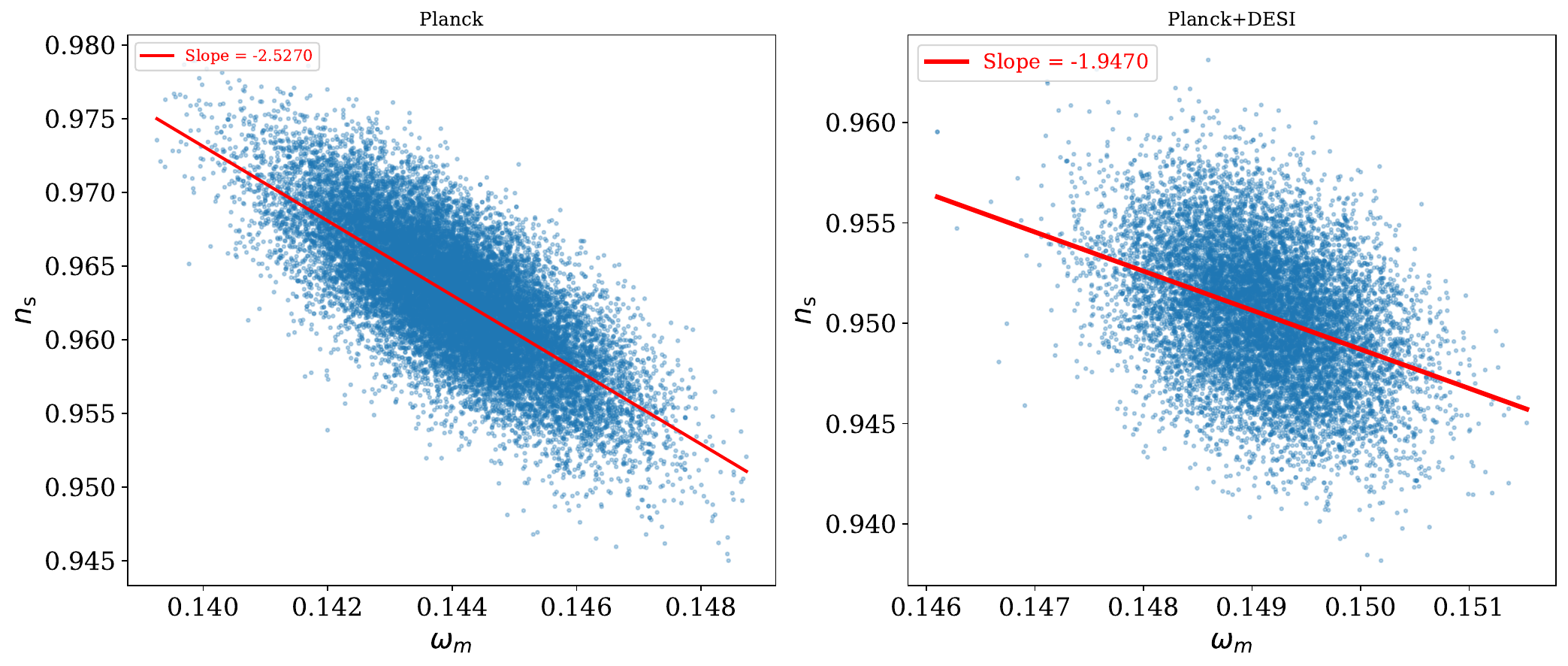}
    \caption{The contours of the parameters $(\omega_{m0}, n_s)$ obtained from Planck and Planck+DESI data for $f$CDM model are shown, together with the degeneracy direction. For the Planck data, the degeneracy direction has a slope of $-2.52$, while for the Planck+DESI data the slope becomes $-1.94$. This indicates a strong degeneracy between the two parameters in the Planck data, where the correlation coefficient between them is $\varrho=-0.7$. In contrast, when DESI data are included, the degeneracy becomes weaker, with a correlation coefficient of $\varrho=-0.39$.}
    \label{fig:dgnrt}
\end{figure}

Since the constraint associated with the peak locations must remain satisfied (due to its high precision and nearly model-independent nature), $\theta_s$ is preserved. As a consequence, the change propagates into the second relation, i.e. $\Omega_{m0} h^2$. In the $f(T)$ model, Planck data alone measures $\omega_{m0}=0.1440 \pm 0.0013$, while Planck+DESI measures $\omega_{m0}=0.14909\pm0.00064$ with an increase $\sim 3.5 \sigma$. Due to the degeneracy between $\Omega_{m0} h^2$ and $n_s$, see Fig. \ref{fig:dgnrt}, maintaining a good fit to the CMB peak heights requires a reduction in $n_s$ to compensate for the increase in $\Omega_{m0} h^2$. Similarly, this explains the shift toward higher $n_s$ values when DESI is included in $\Lambda$CDM, see also~\cite{Ferreira:2025lrd}, where a reduction in $\Omega_{m0} h^2$ requires an increase in $n_s$ in order to maintain a good fit to the data.

The measurement of the spectral index $n_s$ plays a vital role in viable inflationary models. It has been shown that there is a discrepancy between measurements of the CMB and BAO, highlighting the role of dataset consistency in the inferred value of $n_s$~\cite{Ferreira:2025lrd}. Consequently, not only the inferred value of $n_s$ shows the presence of the BAO-CMB tension, but also other parameters are expected to show some levels of tensions due to degeneracies among cosmological parameters~\cite{Efstathiou:1998xx} (see also~\cite{Allali:2025wwi,Allali:2025yvp}). More detailed results are given in Appendix \ref{App:1}.

%%%%%%%%%%%%%%%%%%%%%%%%%%%%%%%%%%%%%%%%%%%%%%%%%%%%
\subsection{Matter power spectrum and large \texorpdfstring{$\sigma_8$}{}}\label{sec:sigma8}
%%%%%%%%%%%%%%%%%%%%%%%%%%%%%%%%%%%%%%%%%%%%%%%%%%%%

\begin{figure}
    \centering
        \includegraphics[width=0.5\linewidth]{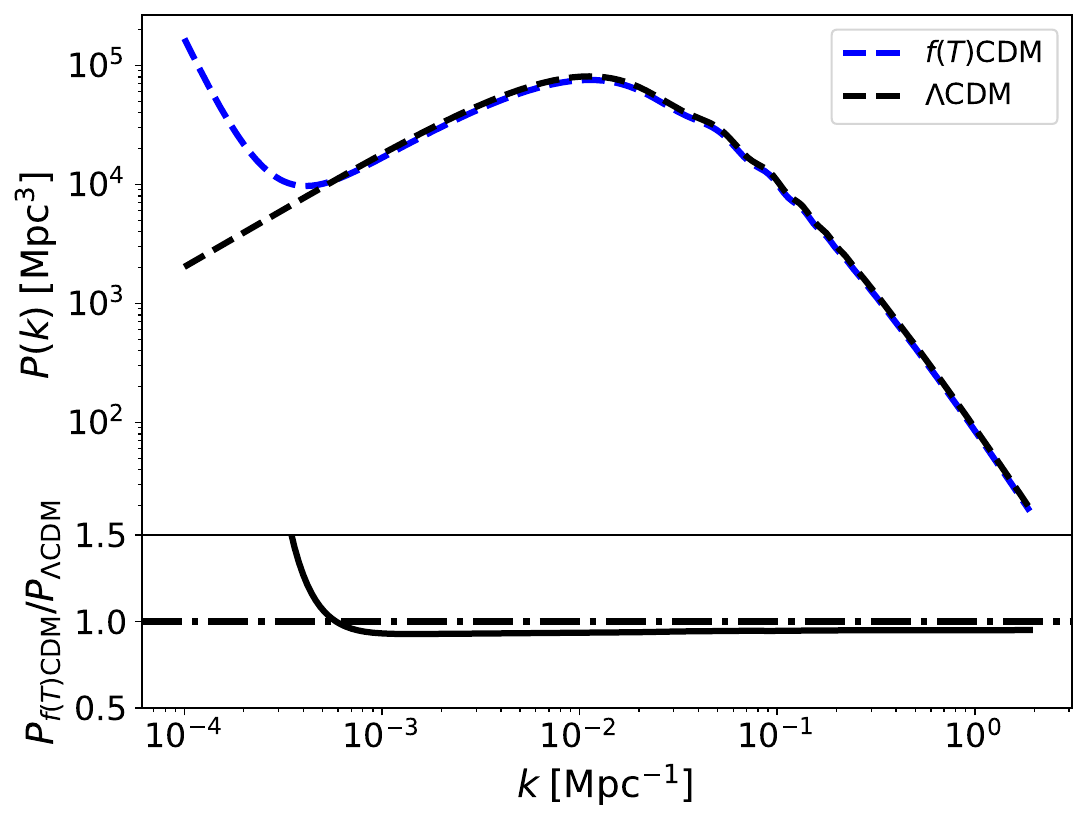}
        \caption{The matter power spectrum for $\Lambda$CDM (dashed curve) and $f$CDM, using the parameters of both models constrained by the Planck data. The deviation in the $f(T)$ matter power spectrum from $\Lambda$CDM becomes significant near the horizon scales $k \sim 10^{-4}$. This is due to large gravitational slip in $f(T)$ gravity at these scales.}
        \label{fig:mttrpwr}
\end{figure}

We evaluate the matter power spectrum for $f$CDM and $\Lambda$CDM using the parameters of both models constrained by the Planck data as shown in Figure \ref{fig:mttrpwr}. Obviously, the $f(T)$ power spectrum at very large scales, i.e. $k<0.001$, becomes larger than $\Lambda$CDM and it becomes very large near the horizon scales where gravitational slip becomes significant, for more details see~\cite{Hashim:2021pkq}. For the modes $k\gtrsim 0.001$, the $\Lambda$CDM power spectrum is larger than that of the $f$CDM model across these scales. This implies that nearly all modes entering the horizon during the matter- and radiation-dominated eras have higher amplitudes in the $\Lambda$CDM matter power spectrum. At first glance, this may appear counterintuitive when compared to the results in Tables \ref{tab:LCDM} and \ref{tab:f1}, since the matter power spectrum in $\Lambda$CDM is larger than in the $f(T)$ model over a wide range of scales, while the inferred value of $\sigma_8$ is higher in the $f(T)$ scenario. In the following, we clarify this apparent contradiction and show that it arises from the interplay between scale dependence and the definition of $\sigma_8$.

The variance of the matter density fluctuations is defined via $\sigma_8$, which represents the amplitude at the scale $R = 8\, h^{-1}\mathrm{Mpc}$
\begin{equation}\label{eq:sgm8}
    \sigma_8^{2} = \frac{1}{2\pi^2} \int_0^\infty dk\, k^{2} P(k) |W(kR)|^2.
\end{equation}
The linear matter power spectrum $P(k, a)$ at a given scale $k$ and scale factor $a$ is governed by the following proportionality
\begin{equation}
    P(k, a) \propto P_{\mathcal{R}}(k) \cdot T^2(k) \cdot D^2(a),
\end{equation}
where $P_{\mathcal{R}}(k)$ is the primordial power spectrum from inflation, $T^2(k)$ is the transfer function and $D^2(a)$ is the linear growth factor. Using a simplified approximation, the linear matter power spectrum $P(k)$ behaves as
\begin{equation}
    P(k) \sim
    \begin{cases}
        k^{n_s}, & k < k_{\mathrm{eq}} \\
        k^{n_s - 4}, & k > k_{\mathrm{eq}}
    \end{cases}
\end{equation}
where $k_{\mathrm{eq}}$ is the scale entering the horizon at matter-radiation equality. The Fourier transform of the spherical top-hat window function is given by
\begin{equation}
    W(kR) = 3 \left( \frac{\sin(kR) - (kR)\cos(kR)}{(kR)^3} \right).
\end{equation}
Since the $f(T)$ model maintains baseline parameter values from Planck data that are nearly identical to those of $\Lambda$CDM, the primordial power spectrum $P_{\mathcal{R}}(k)$ is effectively the same in both models. Therefore, the difference in the inferred value of $\sigma_8$ arises from the net balance of the following two competing effects:

\textit{The first effect} is related to the larger value of the Hubble parameter $H_0$ in $f(T)$ model.
Since $\sigma_8$ is defined as the variance of matter fluctuations smoothed on a scale
\begin{equation}\label{eq:Rscale}
    R = \frac{8}{h}\,\mathrm{Mpc},
\end{equation}
a larger value of $h$ corresponds to a smaller smoothing scale $R$. The window function entering the definition of $\sigma_8$ selects a range of wavenumbers centered around $k \sim 1/R$. Therefore, the decrease in $R$ shifts the dominant contribution of the integral toward larger wavenumbers. As the matter power spectrum generally increases with $k$ for $k < k_{\rm eq}$, this shift enhances the value of $\sigma_8$ in the $f(T)$ model. We note that this conclusion should be generic for any model which gives larger $H_0$.

\begin{figure}
    \centering
    \begin{subfigure}{0.48\linewidth}
    \centering
        \includegraphics[width=\linewidth]{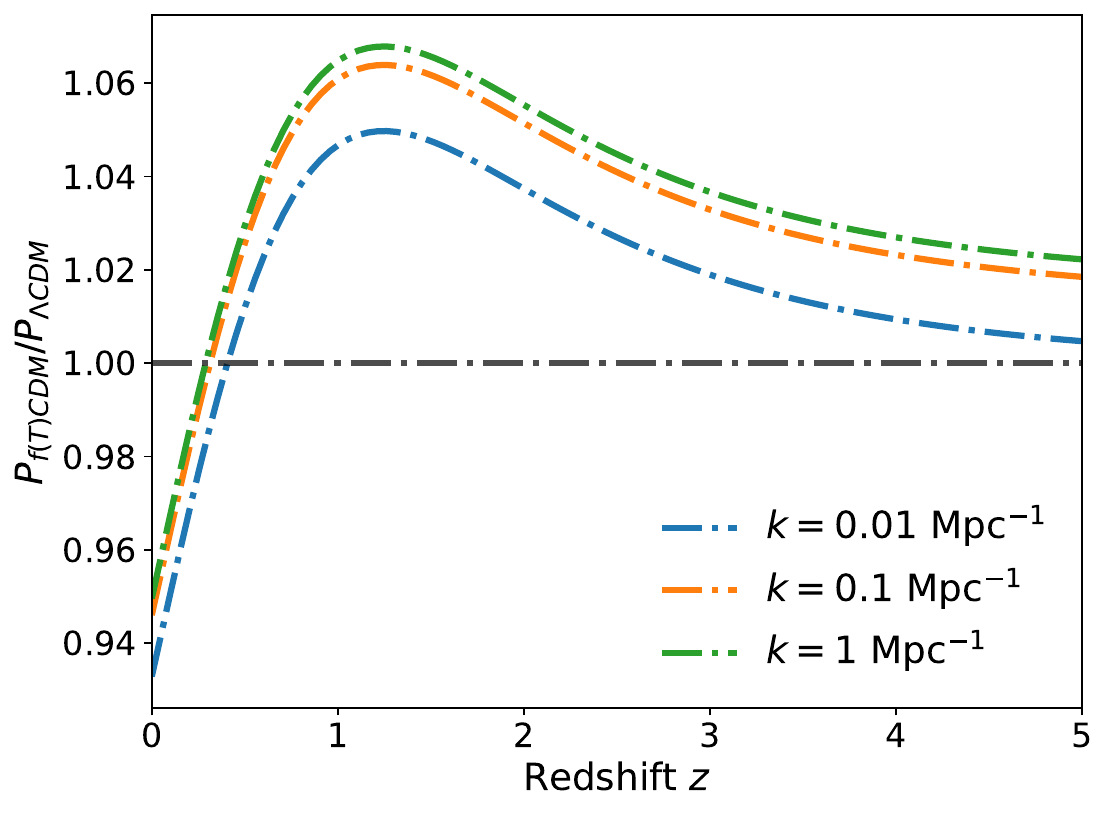}
        \caption{Power spectrum ratio.}
        \label{fig:pk_rto}
    \end{subfigure}
    %\hspace{0.5cm}
    \begin{subfigure}{0.48\linewidth}
    \centering
         \includegraphics[width=\linewidth]{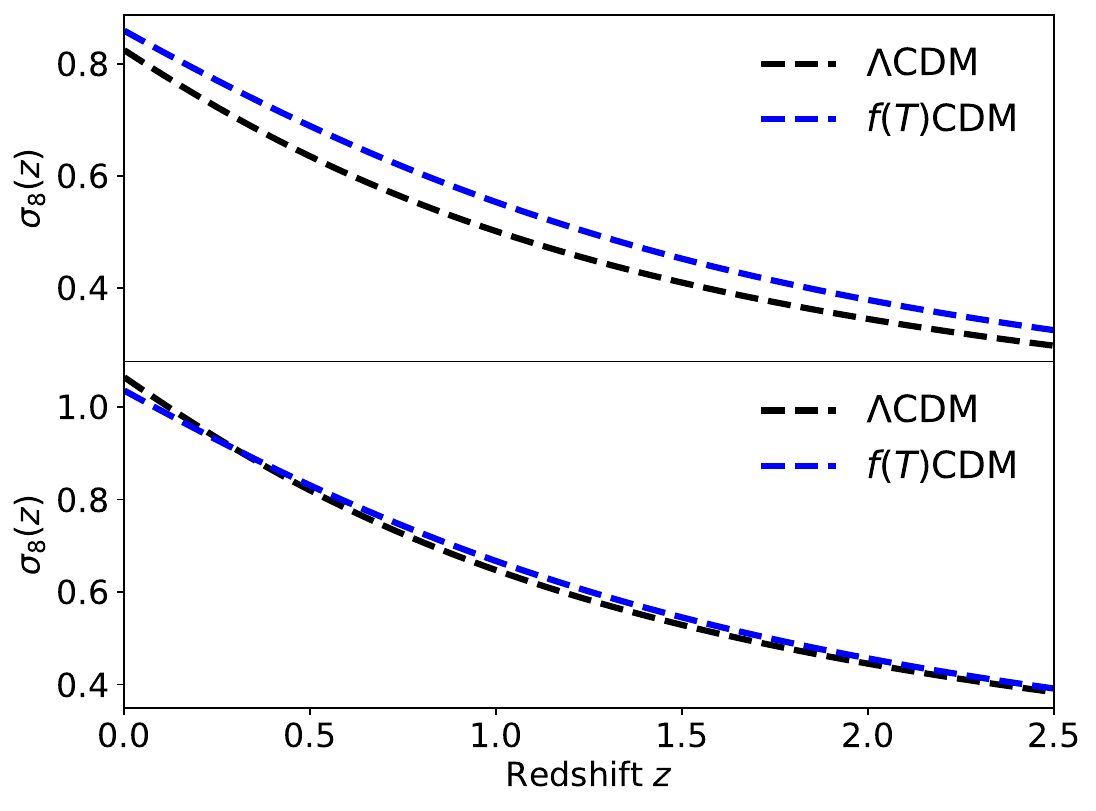}
        \caption{Evolution of $\sigma_8$.}
        \label{fig:sgm8}
    \end{subfigure}
    \caption{ (\subref{fig:pk_rto}) The ratio of the matter power spectrum, $P_{f(T)}/P_{\Lambda \text{CDM}}$, for various scales $k = \{0.01, 0.1, 1\}$ Mpc$^{-1}$ across the redshift range $0 < z < 5$.  All cosmological parameters used in this comparison are consistent with the latest constraints from the \textit{Planck} dataset. (\subref{fig:sgm8}) Comparing the evolution of $\sigma_8$, defined in Eq.~\eqref{eq:sgm8}, is presented for both the $\Lambda$CDM model (dashed black line) and the $f(T)$ model (dashed blue line). For two different units of $k$, the upper plot uses units of $\mathrm{Mpc}^{-1}$ and the lower plot uses $h\,\mathrm{Mpc}^{-1}$. We are using the best-fit values of cosmological parameters estimated from Planck data. }
    \label{fig:full_cosmo_comparison}
\end{figure}

\textit{The second effect} arises from the late--time growth of matter density perturbations, where $\sigma_8$ is proportional to the present--day amplitude of matter fluctuations, i.e. $\sigma_8 \propto \delta_m(z=0)$. Within $f(T)$ teleparallel gravity, in particular, the modified growth equation for matter perturbations, at Newtonian sub-horizon scales, is given by
\begin{equation}\label{eq:delta_evol}
  \ddot{\delta}_m + 2H \dot{\delta}_m - 4\pi GQ \rho_m \, \delta_m = 0.
\end{equation}
At the $Q\to 1$, the GR limit is recovered~\cite{Peebles1980}. Clearly the growth rate is determined by the friction term $2H \dot{\delta}_m$ as well as the modified gravity via the product $QG$ which serves as an effective gravitational constant. In general, faster expansion (higher $H(z)$) suppresses the formation of structures $\delta_m(z)$, while stronger gravity, i.e. $Q > 1$, accelerates the formation. The combined effect of the two factors are scale free in $f(T)$ modified gravity, so they determine the growth at different times.

In our case, it can be shown that $Q>1$ for the present $f(T)$ model at all redshifts which enhances the modified gravity indicating faster growth in $\delta_m$. However, the expansion rate $H(z)$ in the present model varies with time relative to $\Lambda$CDM. Since $H(z)$ is larger in the present $f(T)$ model than in $\Lambda$CDM within the range $0 < z < z_c$, where $z_c$ the redshift at which $H_{f}=H_{\Lambda}$ as seen in Fig. \ref{fig:hbblcmprs}(\subref{fig:fTvLam}), this term acts to suppress the growth in the $f$CDM model in comparison to $\Lambda$CDM at present time as indicated by Eq.~\eqref{eq:delta_evol}. This is on the contrary to the redshift range $z>z_c\sim 0.5$, where a lower $H(z)$ (relative to $\Lambda$CDM) results in faster growth in $\delta_m$.

To verify this claim, we plot the ratio of the matter power spectrum, $P_{f(T)}/P_{\Lambda \text{CDM}}$, see Fig. \ref{fig:full_cosmo_comparison}(\subref{fig:pk_rto}), for several scales $k = \{0.01, 0.1, 1\}$ Mpc$^{-1}$ across the redshift range $0 < z < 5$. The $f$CDM model initially exhibits an enhanced power spectrum at higher redshifts $z>z_c\sim 0.5$, whereas this ratio declines and eventually becomes smaller than unity when universe evolves toward lower redshifts. In this sense, we find that this behaviour is driven by the expansion rate in the $f$CDM model at late times, which suppresses the growth of matter perturbations $\delta_m$ compared to the standard $\Lambda$CDM scenario and consequently $\sigma_8$. However, the net outcome of these two opposing effects results in a larger $\sigma_8$ for the $f(T)$ model compared to $\Lambda$CDM when using Planck data.

According to the above discussion, we note that the apparent discrepancy mentioned at the beginning of this section is justified. In summary, although $\Lambda$CDM exhibits a physically higher power spectrum amplitude than $f$CDM, this does not necessarily imply a larger $\sigma_8$ value. Recalling Eq.~\eqref{eq:Rscale}, we find that $\sigma_8$ is inherently scale-dependent, relying on the smoothing scale $R$ which depends, in turn, on the value of $H_0$. Given that $H_0$ is significantly larger in the $f(T)$ model, we are effectively calculating $\sigma_8$ over two different physical scales for the two models
\[
R_{\Lambda} = \frac{8}{0.67}~\text{Mpc}, \qquad R_{f(T)} = \frac{8}{0.74}~\text{Mpc}.
\]
So, we can say  the reason why $\sigma_8$ is larger in the $f$CDM model from Planck data is not due to an enhancement of growth relative to $\Lambda$CDM. Instead, as it is because $H_0$ is larger. In Fig.~\ref{fig:full_cosmo_comparison}(\subref{fig:sgm8}), we plot the evolution of the mass fluctuation $\sigma(R, z)$ for a scale of $R = 8$ in both $\mathrm{Mpc}$ and $h^{-1}\mathrm{Mpc}$ units. This analysis aims to verify the claim that the larger $\sigma_8$ value in the $f$CDM model is a direct consequence of the different physical scales involved. The comparison clearly demonstrates that in units of $\mathrm{Mpc}^{-1}$ where the $R$ values in both models are identical the $\Lambda$CDM model has a larger $\sigma_8$ than in $f$CDM model at late universe, and consequently $\Lambda$CDM exhibits a larger growth evolution. However, for units of $h\,\mathrm{Mpc}^{-1}$, we see that for all values of $z$, the $\sigma_8$ in $f$CDM becomes greater in comparison to the corresponding value in $\Lambda$CDM; this is due to the larger value of $H_0$ in the $f$CDM model. In this sense, the real physical difference can be better seen in the parameter $S_8$. This because $S_8$ contains $\Omega_{m0}$ which is different in the two models and is inversely proportional to $H_0$ under the condition of fixed matter density, i.e. $\Omega_{m0}h^2$ is fixed.

By combining background data (DESI and supernova) with Planck, a slight tension emerges in baseline parameters like $n_s$ and $\omega_{m0}$ in $f(T)$ gravity compared to $\Lambda$CDM. This introduces an extra effect a shift in the primordial power spectrum to the previously discussed effects. The cumulative impact of these three factors results in an overall larger $\sigma_8$ for the $f(T)$ model with these data sets.

%%%%%%%%%%%%%%%%%%%%%%%%%%%%%%%%%%%%%%%%%%%%%%%%%%%%%%%%%%%%%%%%%%%%%%%%%%%%%%%%%%%%%%%%%%
\section{Summary and Conclusion} \label{sec:Conclu}
%%%%%%%%%%%%%%%%%%%%%%%%%%%%%%%%%%%%%%%%%%%%%%%%%%%%%%%%%%%%%%%%%%%%%%%%%%%%%%%%%%%%%%%%%%
In this work, we have developed a novel and non-trivial infrared (IR) gravity model within the framework of a six-parameter $f(T)$ teleparallel gravity. This development is directly motivated by the exponential IR model originally proposed in~\cite{Awad:2017yod}. By refining the functional form of the torsion scalar, our proposed model effectively recovers the standard $\Lambda$CDM dynamics during the early-time evolution while exhibiting distinct behavior in the late-time universe. A key advantage of this framework is that it maintains the same number of free cosmological parameters as the $\Lambda$CDM model. Therefore, the statistical comparison between our model and the standard cosmological constant baseline is remarkably straightforward and consistent.

The most distinctive feature of the present model is the dynamical evolution of the effective dark energy density, which exhibits a sign-switching behavior starting with a negative density at high redshifts and smoothly transitioning to a positive regime at $z_{tr} \approx 1.6$. This sign-switching behavior provides the necessary flexibility in the expansion history to address current cosmological tensions directly. The observed sign-switching behavior is supported by recent observational evidence. Specifically, this feature was first introduced to resolve the tension between Ly-$\alpha$ observations and the $\Lambda$CDM model~\cite{BOSS:2014hhw}. Furthermore, recent reconstructions using DESI data~\cite{DESI:2024aqx} have further confirmed the preference for such dynamics. Beyond these, similar models, but at higher redshift transitions, may help alleviating the tensions associated with the James Webb Space Telescope (JWST) data, as it produces a larger growth rate than $\Lambda$CDM at early times~\cite{Menci:2024rbq,Menci:2022wia,Forconi:2023hsj,Biagetti:2022ode}, consistent with the formation of massive early galaxies.

The obtained results show that the proposed model yields a high Hubble constant value of $H_0 = 74.48 \pm 0.61~{\rm km}\,{\rm s}^{-1}\,{\rm Mpc}^{-1}$ when using Planck data, effectively resolving the tension with local measurements. While this value decreases to $H_0 \approx 72~{\rm km}\,{\rm s}^{-1}\,{\rm Mpc}^{-1}$ upon the inclusion of late-time data (DESI and Supernovae), the tension remains significantly lower than that of the $\Lambda$CDM model. Upon combining Planck and DESI data, our model exhibits a parameter tension of approximately $2\sigma$ in $n_s$ and $3.5\sigma$ in $\omega_{m0}$ mainly due to $\omega_{c0}$. Notably, while the standard $\Lambda$CDM model also shows shifts in these parameters ($1.2\sigma$ in $n_s$ and a $1.6\sigma$ in $\omega_{m0}$), the directions of these shifts are notably opposite to those observed in the present $f(T)$ teleparallel gravity model.

We remark that the present $f(T)$ model is strongly disfavored in comparison to $\Lambda$CDM in terms of the values of $\chi^2$ of their best fit parameters. We argue that any cosmological model that seeks to resolve the $H_0$ tension via late universe modifications will inevitably encounter similar behaviour when combined with late-time DESI observations. This suggests a combined modification at early and late universe evolution may resolve the Hubble tension where the shifts in other cosmological observables can be compensated adequately.

%\clearpage

%%%%%%%%%%%%%%%%%%%%%%%%%%%%%%%%%%%%%%%%%%%%%%%%%%%%%%%%%%%%%%%%%%%%%%%%%%%%%%%
\section*{Acknowledgements}
M.H. and W.E. would like to thank Amr El-Zant for the discussion about the negative dark energy density and its role to address tensions such as the $H_0$-tension. Also, the authors would like to thank Eleonora Di Valentino for several discussions about the results of this work. This article is based upon work from COST Action CA21136 \emph{Addressing observational tensions in cosmology with systematics and fundamental physics} (CosmoVerse), supported by COST (European Cooperation in Science and Technology). JLS would also like to acknowledge funding from ``Xjenza Malta'' as part of the ``Technology Development Programme'' DTP-2024-014 (Cosmic Learning) Project.
%%%%%%%%%%%%%%%%%%%%%%%%%%%%%%%%%%%%%%%%%%%%%%%%%%%%%%%%%%%%%%%%%%%%%%%%%%%%%%%
\appendix
\section{CMB-BAO tensions}\label{App:1}
\begin{figure}[h!]
    \centering
    \includegraphics[width=0.85\linewidth]{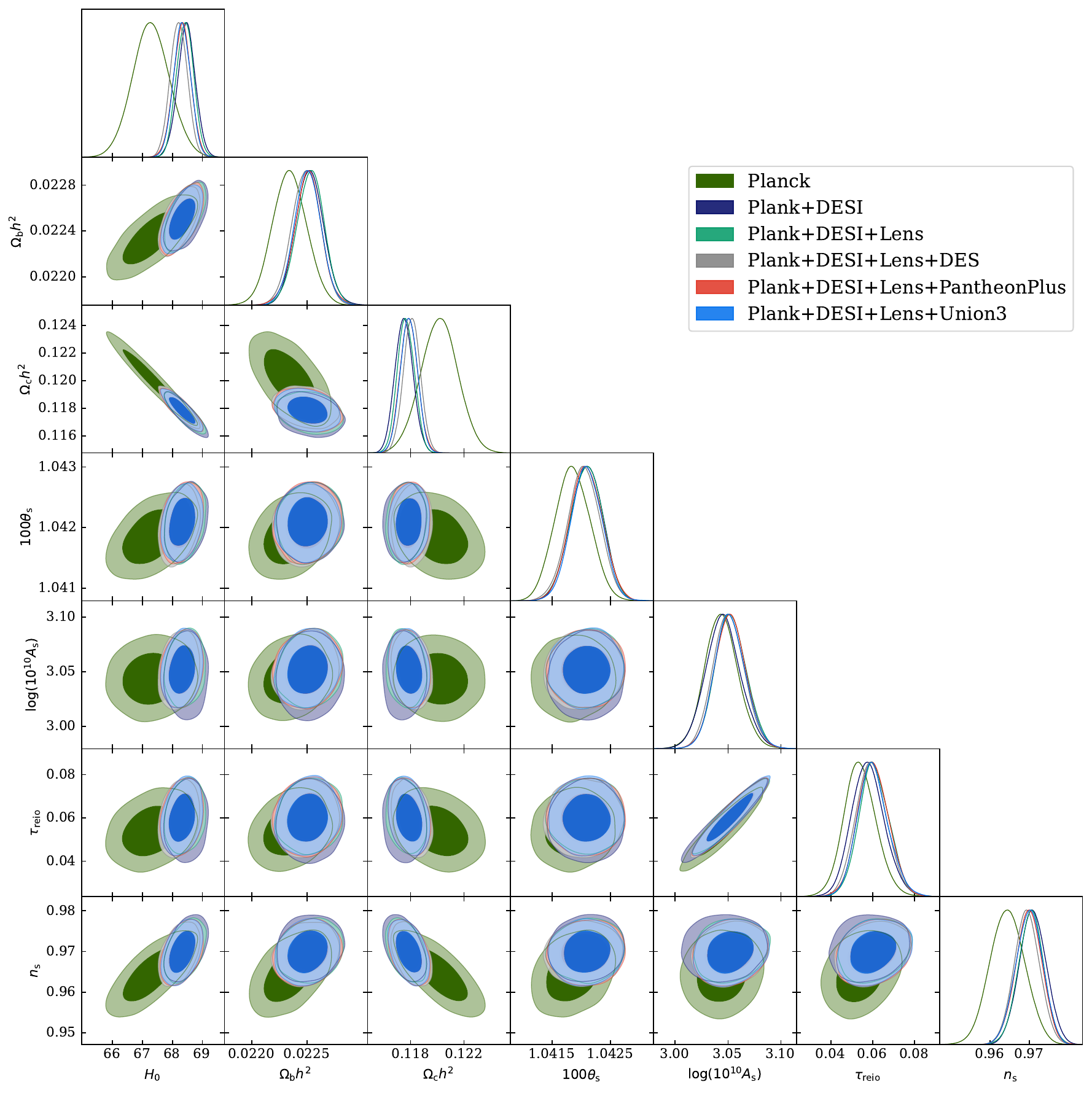}
    \caption{The 68\% (dark color) and 95\% (faint color) confidence level (CL) contours of the cosmological parameters for the $\Lambda$CDM  model using all data set We also include the one-dimensional posterior distributions for the parameters.}
    \label{fig:LCDM_all}
\end{figure}

\begin{figure}[h!]
    \centering
    \includegraphics[width=0.9\linewidth]{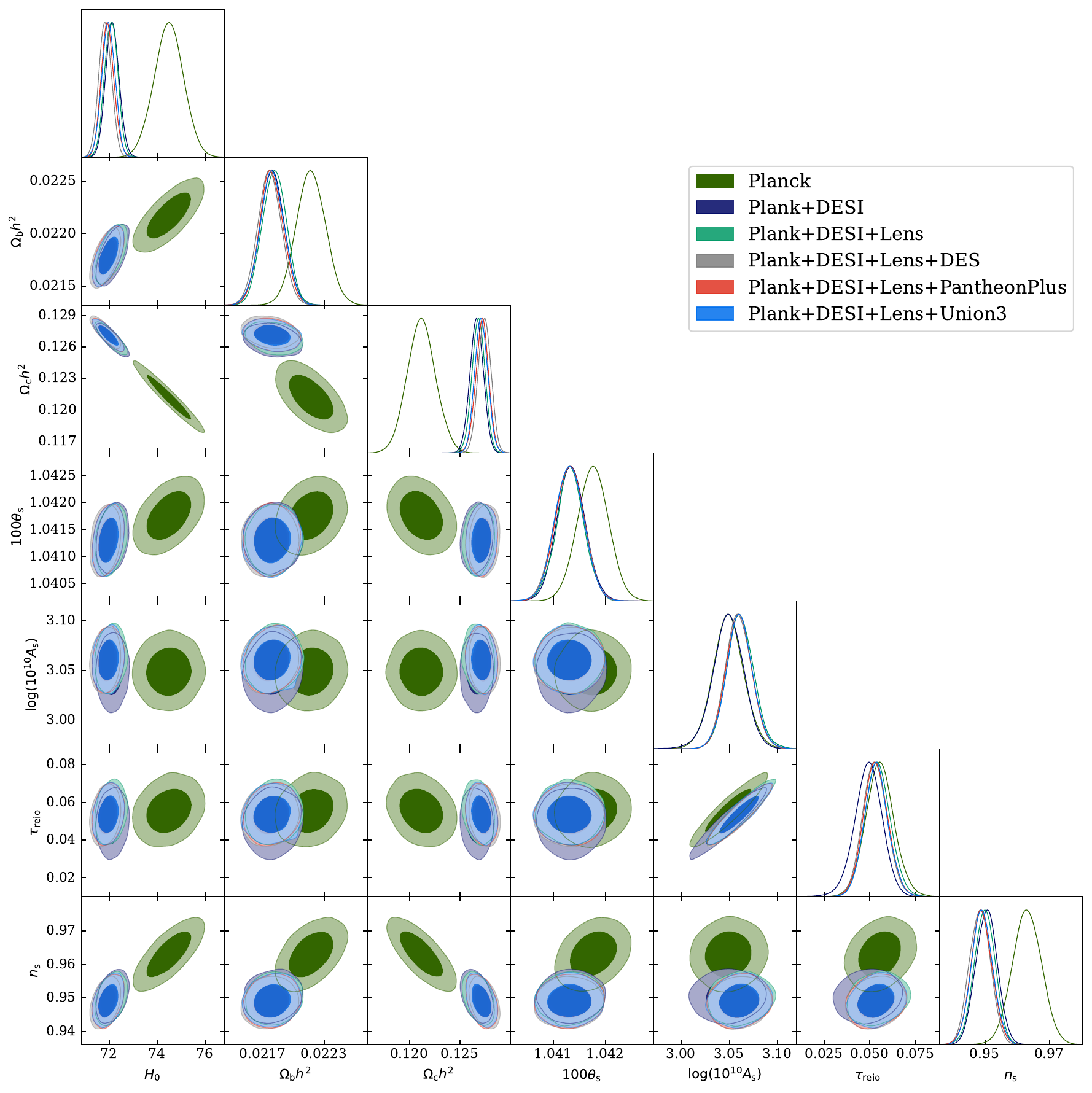}
    \caption{The 68\% (dark color) and 95\% (faint color) confidence level (CL) contours of the cosmological parameters for the $f$CDM  model using all data set We also include the one-dimensional posterior distributions for the parameters.}
    \label{fig:f(T)_all}
\end{figure}

The spectral index $n_s$ tension has been discussed earlier in Subsection \ref{sec:nstension}. Indeed, the measurement of $n_s$ plays a vital role in viable inflationary models. It has been shown that there is a discrepancy between measurements of the CMB and BAO, highlighting the role of dataset consistency in the inferred value of $n_s$ in agreement with~\cite{Ferreira:2025lrd}. Consequently, not only the inferred value of $n_s$ shows the presence of the BAO-CMB tension, but also other parameters are expected to show this tension due to degeneracies among cosmological parameters~\cite{Efstathiou:1998xx} (see also~\cite{Allali:2025wwi,Allali:2025yvp}). In Figures \ref{fig:LCDM_all} and \ref{fig:f(T)_all}, we give the 2D contour plots of the six parameters in addition to the $H_0$ value for $\Lambda$CDM and $f$CDM, respectively. These plots show clearly the tensions, among cosmological parameters, between Planck-CMB alone verses Planck+DESI and also other datasets for each model. Remarkably, the plots show that the $f$CDM model introduces new degeneracy directions opposite to $\Lambda$CDM.

%\clearpage
%%%%%%%%%%%%%%%%%%%%%%%%%%%%%%%%%%%%%%%%%%%%%%
%\bibliographystyle{ieeetr}
\bibliographystyle{apsrev4-1}
\bibliography{Ref}
%%%%%%%%%%%%%%%%%%%%%%%%%%%%%%%%%%%%%%%%%%%%%%

\end{document}